%% file: main.tex
\documentclass[10pt,epsfig]{article} 
\usepackage{tikz}
\usepackage{longtable}
\usepackage{enumerate}
\usepackage{float}
\usepackage{subfig}
\usepackage{color}
\usepackage{xcolor}
\usepackage{slashed}
\usepackage{multirow}
\usepackage{cite}

\let\counterwithin\relax
\usepackage{chngcntr}
\usepackage{amssymb, amsmath,mathrsfs}

\usepackage{graphics}
\usepackage{graphicx}
\usepackage{epsf}
\usepackage{epsfig}
\usepackage{float}
\usepackage{makecell}
\usepackage{multirow}
\usepackage{color}
\usepackage{xcolor}
\usepackage{simplewick}

\usepackage[utf8]{inputenc}
\usepackage{amsmath}
\usepackage{amssymb}
\usepackage{subfig}
\usepackage[normalem]{ulem}

\usepackage{mathtools}
\usepackage{amsmath}
\usepackage{diagbox}

\newcommand\undermat[2]{
	\makebox[0.5pt][l]{$\smash{\underbrace{\phantom{%
					\begin{matrix}#2\end{matrix}}}_{ \let\scriptstyle\textstyle\text{\large $#1$}}}$}#2}
\newcommand\overmat[2]{
	\makebox[-1pt][l]{$\smash{\overbrace{\phantom{%
					\begin{matrix}#2\end{matrix}}}^{ \let\scriptstyle\textstyle\text{\large $#1$}}}$}#2}    
\usepackage{tikz}
\usepackage[vcentermath]{youngtab}
\usepackage{slashed} 
\usepackage[font=small]{caption} 
\usepackage{times} 

\usepackage{bm}       
\usepackage{bbm} 

\usepackage{relsize}  

\usepackage{autobreak}  

\usepackage[colorlinks=true,
linkcolor=blue,
urlcolor=red,
citecolor=red]{hyperref}

\usepackage{makeidx} 
\usepackage{bbding} 
\usepackage{listings} 
\usepackage{ytableau}
\usepackage[utf8]{inputenc}
\usepackage[T1]{fontenc}
\usepackage{lmodern}

\ytableausetup
{boxsize=1.0em}

\long\def\rpl#1!!#2!!{\textcolor{red}{#1} \textcolor{blue}{#2}}

\usepackage[top=1in, left=0.95in, bottom=1.1in, right=0.95in]{geometry}
\def\baselinestretch{1.27}
\usepackage[toc]{appendix}

\newcommand{\comments}[1]{{}}

\newcommand{\beq}{\begin {equation}}
\newcommand{\eeq}{\end   {equation}}
\newcommand{\bea}{\begin {eqnarray}}
\newcommand{\eea}{\end   {eqnarray}}
\newcommand{\beqa}{\begin {eqnarray}}
\newcommand{\eeqa}{\end   {eqnarray}}
\newcommand{\baa}{\begin {array}   }
\newcommand{\eaa}{\end   {array}   }
\newcommand{\bit}{\begin {itemize} }
\newcommand{\eit}{\end   {itemize} }
\newcommand{\be }{\begin {equation}}
\newcommand{\ee }{\end   {equation}}

\newcommand{\bft}{\mathbf{T}}
\newcommand{\bfu}{\mathbf{U}}
\newcommand{\bfv}{\mathbf{V}}

\newcommand{\calo}{\mathcal{O}}
\newcommand{\caly}{\mathcal{Y}}
\newcommand{\calm}{\mathcal{M}}
\newcommand{\calb}{\mathcal{B}}

\newcommand{\lra}[1]{\langle #1 \rangle}
\newcommand{\lrs}[1]{[ #1 ]}
\newcommand{\ione}{i_1}
\newcommand{\itwo}{i_2}
\newcommand{\jone}{j_1}
\newcommand{\jtwo}{j_2}
\newcommand{\ithree}{i_3}

\newcommand{\ifour}{i_4}
\newcommand{\jfour}{j_4}
\newcommand{\ifive}{i_5}
\newcommand{\jfive}{j_5}
\newcommand{\isix}{i_6}
\newcommand{\jsix}{j_6}
\newcommand{\iseven}{i_7}
\newcommand{\jseven}{j_7}

\newcommand{\cthree}{c_3}
\newcommand{\cfour}{c_4}
\newcommand{\aone}{a_1}
\newcommand{\atwo}{a_2}

\newcommand{\bthree}{b_3}

\newcommand{\bfour}{b_4}

\begin{document}

\begin{center}

{\Large \textbf  {Complete NNLO Operator Bases in Higgs Effective Field Theory}}\\[10mm]

Hao Sun$^{a, b}$\footnote{sunhao@itp.ac.cn}, Ming-Lei Xiao$^{c, d}$\footnote{minglei.xiao@northwestern.edu}, Jiang-Hao Yu$^{a, b, e, f, g}$\footnote{jhyu@itp.ac.cn}\\[10mm]

\noindent 
$^a${\em \small CAS Key Laboratory of Theoretical Physics, Institute of Theoretical Physics, Chinese Academy of Sciences,    \\ Beijing 100190, P. R. China}  \\
$^b${\em \small School of Physical Sciences, University of Chinese Academy of Sciences,   Beijing 100049, P.R. China}   \\
$^c${\em \small Department of Physics and Astronomy, Northwestern University, Evanston, Illinois 60208, USA}\\
$^d${\em \small High Energy Physics Division, Argonne National Laboratory, Lemont, Illinois 60439, USA}\\
$^e${\em \small Center for High Energy Physics, Peking University, Beijing 100871, China} \\
$^f${\em \small School of Fundamental Physics and Mathematical Sciences, Hangzhou Institute for Advanced Study, UCAS, Hangzhou 310024, China} \\
$^g${\em \small International Centre for Theoretical Physics Asia-Pacific, Beijing/Hangzhou, China}\\[10mm]

\date{\today}   
          
\end{center}

\begin{abstract}

For the first time, we list the complete and independent set of operators at the next-to-next-to-leading order (NNLO) in the Higgs effective field theory (HEFT). The Young tensor technique utilized in this work guarantees the completeness and independence of the on-shell amplitude basis while the Adler zero condition imposes non-linear symmetry on the Nambu-Goldstone bosons that play the central role in the chiral Lagrangian.
The spurion fields are incorporated into the gauge structure of operators following the Littlewood-Richardson rule to accommodate custodial symmetry breaking. 
We enumerate 11506 (1927574) NNLO operators for one (three) flavor of fermions for the electroweak chiral Lagrangian with the light Higgs. Below the electroweak symmetry breaking scale, the dimension-8 standard model effective field theory (SMEFT) operators could be matched to these HEFT operators at both the NLO and NNLO orders. 


\end{abstract}

\newpage

\setcounter{tocdepth}{3}
\setcounter{secnumdepth}{3}

\tableofcontents

\setcounter{footnote}{0}

\def\baselinestretch{1.5}
\counterwithin{equation}{section}

\newpage

\section{Introduction}
\label{sec:intro}
In the last several decades, the standard model (SM) of particle physics has shown its grand validity. However, the failure of discovering particles beyond the SM at the Large Hadron Collider (LHC) implies that there is a considerable energy gap between the SM particles and new physics. Due to the scale separation, various new physics effects below the energy threshold of new physics particle can be characterised by the effective field theory (EFT) framework. Pioneered by Weinberg~\cite{Weinberg:1978kz}, EFT has been developed to be a systematic framework to parametrize the underlying physics at low energy scale. In such bottom-up approach, the building blocks at the low energy are used to construct all the operators satisfying the specific symmetries with proper power-counting. Therefore, all the operators can be organised order by order, and in each order, the Lagrangian is the linear combination of all the independent operators, and the coefficient of each operator is called the Wilson coefficient, carrying information from the underlying dynamics at the high energy. 

Adopting the SM particles as the building blocks, and imposing the gauge symmetry in the SM, the higher dimensional operators can be constructed and organized order by order in terms of the canonical dimension powers. The constructed EFT is the standard model effective field theory (SMEFT). Since the dimension-5 operators were presented by Weinberg~\cite{Weinberg:1979sa}, many progresses on the operator bases have been made~\cite{ Buchmuller:1985jz, Grzadkowski:2010es, Lehman:2014jma, Liao:2016hru, Li:2020gnx, Murphy:2020rsh, Li:2020xlh, Liao:2020jmn,Li:2020zfq,Henning:2015alf}. A general algorithm, implemented in a Mathematica package ABC4EFT~\cite{Li:2022tec}, has been proposed to construct the independent and complete SMEFT operator bases up to any mass dimension. 



Nowadays, the dimension-8 operators of the SMEFT~\cite{Li:2020gnx, Murphy:2020rsh} receive more and more attention theoretically and experimentally. Below the electroweak (EW) scale, the SM gauge symmetry is broken down to $SU(3) \times U(1)_{\rm em}$, along with the Higgs doublet broken down to singlet. Thus below the EW scale the SMEFT is not suitable to describe new physics effects any more. On the other hand, due to the approximated custodial symmetry in the Higgs sector, the EFT can be characterised by the Coleman-Callan-Wess-Zumino (CCWZ) formalism~\cite{Coleman:1969sm,Callan:1969sn}, known as the electroweak chiral Lagrangian with the light Higgs boson (H-EWChL), or the Higgs effective field theory (HEFT), see Ref.~\cite{Appelquist:1980vg,Longhitano:1980iz,Longhitano:1980tm,Feruglio:1992wf,Herrero:1993nc,Herrero:1994iu} for early developments and Ref.~\cite{Buchalla:2012qq,Alonso:2012px,Buchalla:2013rka,Brivio:2013pma,Pich:2015kwa,Pich:2016lew,Brivio:2016fzo,Merlo:2016prs,Krause:2018cwe} after the Higgs discovery.  The HEFT provides a more general realization of the EW dynamics, which includes the SMEFT as a particular case~\cite{Falkowski:2019tft,Agrawal:2019bpm,Cohen:2020xca}. The HEFT Lagrangian has been constructed up to next-to-leading order (NLO), including the fermion sector~\cite{Buchalla:2012qq,Alonso:2012px,Buchalla:2013rka,Brivio:2013pma,Pich:2015kwa,Pich:2016lew,Brivio:2016fzo,Merlo:2016prs,Krause:2018cwe}, without considering the flavor structures. Recently the complete and independent NLO operators are presented in Ref.~\cite{Sun:2022ssa}, and the flavor structures of the operators are considered.  


If one would like to match the dimension-8 SMEFT operators to the HEFT operators after the EW symmery breaking, the HEFT Lagrangian should be constructed up to the next-to-next-to leading order (NNLO), to capture various effects only appearing at the dimension-8 SMEFT. For example, to investigate the effective operators contributing to the genuine quartic gauge–boson couplings, the relevant bosonic chiral Lagrangian at $\mathcal{O}(p^6)$ on the quartic gauge couplings has been written in Ref.~\cite{Eboli:2016kko} and the connection to the bosonic dimension-8 SMEFT operators is also discussed. Furthermore, the one loop renormalization of the NLO HEFT operators has been considered~\cite{Guo:2015isa,Alonso:2017tdy,Buchalla:2017jlu,Buchalla:2020kdh}. According to the power counting rules, one-loop renormalization of the NLO operators should be comparable with effects caused by the NNLO HEFT operators.
However, in literature the NNLO operators counted as the order $\mathcal{O}(p^5)$ and $\mathcal{O}(p^6)$ have not yet been constructed systematically.


For the first time, we construct the complete and independent NNLO operators with the flavor structures using Young tensor technique developed in Ref.~\cite{Li:2020xlh,Li:2020gnx,Li:2022tec,Li:2020zfq} with certain improvements on the operators involving in the Nambu-Goldstone Boson (NGB) and the spurion field parametrizing the custodial symmetry breaking:
\begin{itemize}
    \item For the operators involving the NGBs, the Adler zero condition implies that the on-shell amplitudes corresponding to the Lorentz structure of the operators should vanish in the soft limit of the NGB momentum~\cite{Adler:1964um,Adler:1965ga, Low:2014nga, Low:2014oga, Cheung:2014dqa, Cheung:2015ota, Low:2019ynd, Dai:2020cpk,Low:2022iim, Sun:2022ssa}. Thus we need to impose this constraint on the Lorentz basis obtained by the Young tensor method to obtain the reduced Lorentz structures, which are usually of a subspace of the original Lorentz space. 
    
    \item Since the spurions are frozen degrees of freedom, unlike the dynamical fields, the spurion should not enter the Lorentz sector. Instead, it only plays a role in constructing the $SU(2)$ invariant together with other dynamical fields. 
    Furthermore, we should avoid the appearance of self-contracted spurion fields, such as $\delta_{IJ}T^IT^J$, because they are redundant in describing the symmetry breaking pattern.

\end{itemize}
With these improvements on the Young tensor technique, we could obtain that there are 11506 (1927574) NNLO operators with one (three) generations of the SM fermions. 

The paper is organised as follows. In section~\ref{sec:comp}, we briefly review the building blocks and the leading-order Lagrangian of the HEFT, and present the chiral power-counting scheme. In section~\ref{sec:onsh}, we review the Young tensor method to construct the complete and independent effective operators, and focus on the Adler zero condition on the operators involving the NGBs, and present how to deal with the spurion in the Young tensor method in section~\ref{sec:spurion}. 
Based the these, we list the complete NNLO operators of the HEFT in section~\ref{sec:next}, and draw conclusion in section~\ref{sec:conc}. 

\section{Electroweak Chiral Lagrangian}
\label{sec:comp}

The Higgs sector in the SM has a larger global symmetry $SU(2)_L \times SU(2)_R$ than the gauge symmetry of the SM Lagrangian, which is spontaneously broken down to the custodial $SU(2)_C$ symmetry by the Higgs vacuum expectation value (VEV). The coset pattern $\mathcal{G}\rightarrow\mathcal{H}$ with identifying $\mathcal{G}=S U(2)_{L} \times S U(2)_{R}$ and $\mathcal{H}=S U(2)_{C}$ can be described by the non-linearized Nambu-Goldstone Boson fields along with the Higgs singlet using the CCWZ formalism~\cite{Coleman:1969sm,Callan:1969sn}, adding additional explicit breaking terms from the gauge and Yukawa type interactions. In this section, we will briefly review the construction of the leading order (LO) Lagrangian, and discuss how the NLO and NNLO operators are counted based on the chiral power-counting scheme.


\subsection{Building blocks and the LO Lagrangian}
\label{ssec:lead}

The Lagrangian of the Higgs sector in the gaugeless limit ($g, g' \to 0$) reads
\be
\mathcal{L}_{\rm Higgs} = \partial_\mu H^\dagger \partial^\mu H - \lambda \left( H^\dagger H - \frac{v^2}{2} \right)^2\,,
\ee
where $H \equiv (\phi^+, \phi^0)^T$ denotes the $SU(2)_L$ doublet Higgs and the Lagrangian is invariant under the $SU(2)_L$ symmetry. 
In fact, there is another global $SU(2)_R$ symmetry hidden in this Lagrangian if one rewrite the same Lagrangian by introducing another field $H_R \equiv (\phi^0, \phi^-)^T$. This enlarged symmetry $SU(2)_L \times SU(2)_R$ can be made explicit by re-expressing the Higgs field in terms of a bi-fundamental scalar field $\Sigma$ that transforms under the global symmetry
\bea
\Sigma \equiv\left(\begin{array}{cc}\tilde{H} &  H\end{array}\right)=\left(\begin{array}{cc} \phi^{0*} & \phi^{+} \\ -\phi^{-} & \phi^0\end{array}\right) \quad \longrightarrow \quad \mathfrak{g}_L\,\Sigma\, \mathfrak{g}_R^\dagger, \qquad (\mathfrak{g}_L,\mathfrak{g}_R)\in \mathcal{G}\,.
\eea
Then the Lagrangian in the Higgs sector becomes
\be
\label{eq:lh}
\mathcal{L}_{\rm Higgs}  = \frac{1}{2}\lra{\partial_\mu\Sigma^\dagger\partial^\mu\Sigma}-\frac{\lambda}{4} \left(\lra{\Sigma^\dagger\Sigma}-v^2 \right)^2\,,
\ee
where $\lra{\dots}$ represents $SU(2)_L$ matrix trace. 
It is more convenient to parametrize the Goldstone fields in terms of the unitary matrix $\bfu(x) = \exp(\Pi(x)/f)$:
\begin{equation}
    \Sigma(x) \equiv \frac{h(x)+v}{\sqrt{2}}\,\bfu = \frac{h(x)+v}{\sqrt{2}}\, {\exp} \left[\Pi(x)/f\right]\,,
\end{equation}
which separates the NGBs from the Higgs mode. 
The above Goldstone matrix contains three NGBs,
\begin{equation}
    \Pi(x) = \vec{\pi}(x)\cdot\frac{\vec{\sigma}}{2}\,,
\end{equation}
originated from the global symmetry breaking pattern $SU(2)_L \times SU(2)_R/SU(2)_C$. 
Thus the Lagrangian in Eq.~\ref{eq:lh} takes the form that
\begin{equation}
    \mathcal{L}_H = \frac{1}{2}\partial_\mu h\partial^\mu h  + \frac{v^2}{2}\lra{\partial_\mu \bfu \partial^\mu\bfu^\dagger}\,\mathcal{F}(h) - \frac{\lambda}{4}\left(\frac{h^2}{2}+hv-\frac{v^2}{2}\right)^2\,,
\end{equation}
where the $\mathcal{F}$ is dimensionless polynomial
\begin{equation}
    \mathcal{F}(h) = 1+2\frac{h}{v}+\frac{h^2}{v^2}\,.
\end{equation}

Recovering the gauge symmetry would introduce the explicit custodial symmetry breaking. In the above Lagrangian, promoting the group $SU(2)_L$ and the third component of the group $SU(2)_R$ to be local, the ordinary derivatives would be replaced by the covariant derivatives defined as
\be
D_\mu \Sigma = \partial_\mu \Sigma - ig\hat{W}_\mu\Sigma + ig'Y\Sigma \hat{B}_\mu\,, \quad \hat{W}_{\mu\nu}=\vec{W}_{\mu\nu}\cdot\frac{\vec{\sigma}}{2},\quad \hat{B}_{\mu\nu}=B_{\mu\nu}\frac{\sigma_3}{2}\,,
\ee
where $\vec{W}_{\mu\nu},B_{\mu\nu}$ are the gauge fields in the SM. The gauge fields $\hat{W}_{\mu\nu}$ transforms as the triplet of $SU(2)_L$,
\begin{equation}
\hat{W}_{\mu\nu} \rightarrow \mathfrak{g}_L\hat{W}_{\mu\nu} \mathfrak{g}^\dagger_L,\quad \mathfrak{g}_L\in SU(2)_L\,,
\end{equation}
while $\hat{B}_{\mu\nu}$ transforms as $SU(2)_L$ singlet, and thus the explicit custodial symmetry breaking is parametrized by the spurion field ${\mathcal T}_R = \sigma_3/2$ with $\hat{B}_{\mu\nu}=B_{\mu\nu} {\mathcal T}_R $.  

Introducing the SM fermions Yukawa terms would also break the custodial symmetry explicitly. Let us rewrite the SM fermion fields $\psi_{L/R}=P_{L/R}\psi$ that transform covariant under the global symmetry
\bea
Q_L &=& \left(\begin{array}{c}u_L\\d_L\end{array}\right) \quad \rightarrow \quad \mathfrak{g}_{L} Q_L\,, \quad \quad
Q_R = \left(\begin{array}{c}u_R\\d_R\end{array}\right) \quad \rightarrow \quad \mathfrak{g}_{R} Q_R\,, \\ 
L_L &=& \left(\begin{array}{c}\nu_L\\e_L\end{array}\right) \quad \rightarrow \quad \mathfrak{g}_{L} L_L\,, \quad \quad
L_R = \left(\begin{array}{c}\nu_R\\e_R\end{array}\right) \quad \rightarrow \quad \mathfrak{g}_{R} L_R\,,
\eea
Note that the right-handed fermions are the $SU(2)_R$ doublets and thus the $U(1)_Y$ symmetry in the SM is promoted to the $U(1)_X$ symmetry, where $X=(B-L)/2$ is half of the baryon number $B$ minus the lepton number $L$. 
The Yukawa Lagrangian takes more compact form
\bea
	{\mathcal L}_{\rm Yukawa} = - v \overline{\psi_L} \bfu(\Pi) {\mathcal Y}^\psi_R  \psi_R + h.c. \quad {\textrm{with}} \quad {\mathcal Y}^\psi_R  \quad \longrightarrow \quad \mathfrak{g}_{R} {\mathcal Y}^\psi_R \mathfrak{g}^\dagger_{R}\,,
\eea
where $\psi$ takes $Q$ and $L$ and $\mathcal{Y}^\psi_R$ is a $2\times 2$ matrix and takes the form
\bea
{\mathcal Y}_R^Q = \frac{1 }{2}(y_u  + y_d)  + \frac{\sigma_3}{2} (y_u  - y_d), \quad \mathcal{Y}^L_R =
\frac{1 }{2}(y_e  + y_\nu)  + \frac{\sigma_3}{2} (y_e  - y_\nu)\,,\label{eq:yukawa}
\eea
where $y_\nu= 0$ if no right-handed neutrinos. 
Note that the $ \frac{\sigma_3}{2}$ term above parametrizes the custodial symmetry breaking in the Yukawa term. Therefore, the spurion in the fermion sector takes the same form as the one in the gauge sector $\mathcal{T}_R=\sigma_3/2$. 

The above form of the Lagrangian can can also be obtained in the CCWZ formalism \cite{Coleman:1969sm,Callan:1969sn} of the symmetry breaking pattern $\mathcal{G}\rightarrow\mathcal{H}$ with identifying $\mathcal{G}=S U(2)_{L} \times S U(2)_{R}$ and $\mathcal{H}=S U(2)_{V}$, which provides a systematic way to write effective Lagrangian that allow to manifest the symmetries of the theory.
The Goldstone matrix $\bfu$ takes the form $\bfu(x)=\exp(i\Pi(x)/v)$, which transforms under $\mathcal{G}$ as bi-doublet, 
\be
\bfu \rightarrow \mathfrak{g}_L\bfu\mathfrak{g}_R^{\dagger},\quad (\mathfrak{g}_L,\mathfrak{g}_R)\in\mathcal{G}\,.
\ee
Let us collect all the building blocks would appear in the chiral Lagrangian 
\bea
	h,\quad \bfu, \quad\psi_L, \quad\psi_R, \quad \hat{W}_{\mu \nu}, \quad\hat{B}_{\mu \nu}, \quad\hat{G}_{\mu \nu}, \quad {\mathcal T}_R.
\eea
which transforms differently under the global chiral symmetry. 
For the convenience of constructing higher-dimension operators, we can redefine these building blocks with $\bfu$ to make them transform solely under $SU(2)_L$~\cite{Grinstein:2007iv,Buchalla:2012qq,Buchalla:2013rka,Buchalla:2013eza,Gavela:2014vra,Krause:2018cwe,Alonso:2012px,Brivio:2013pma,Brivio:2016fzo,Merlo:2016prs}, 
\bea
	\bfv_\mu(x) = i\bfu(x) D_\mu \bfu(x)^\dagger, \quad  &\longrightarrow & \quad \mathfrak{g}_L \bfv_\mu \mathfrak{g}_L^\dagger \label{eq:vmu}\\
	\hat{W}_{\mu \nu}  \quad  &\longrightarrow & \quad \mathfrak{g}_L \hat{W}_{\mu \nu} \mathfrak{g}_L^\dagger \\
	\hat{B}_{\mu \nu} \quad  &\longrightarrow & \quad  \hat{B}_{\mu \nu} \\
	\hat{G}_{\mu \nu} \quad  &\longrightarrow & \quad  \hat{G}_{\mu \nu} \\
	\mathbf{T} = \bfu \mathcal{T}_{R} \bfu^{\dagger} \quad &\longrightarrow & \quad \mathfrak{g}_{L} \mathbf{T} \mathfrak{g}_{L}^\dagger\\
	\psi_{L}  \quad &\longrightarrow & \quad \mathfrak{g}_{L} \psi_{L} \\ 
	 \bfu \psi_{R} \quad &\longrightarrow & \quad \mathfrak{g}_{L} \bfu \psi_{R}\\ 
	h\quad &\longrightarrow & \quad h
\eea 
We summarise these building blocks and their representations of Lorentz and gauge groups in Tab~\ref{tab:buildingblocks}. There are other redefinition of the building blocks such as the Ref~\cite{Pich:2015kwa,Pich:2016lew,Pich:2018ltt}, and different schemes actually give the same operators set. 

In terms of the building blocks, the LO Lagrangian takes the form that
\begin{align}
\mathcal{L}_2 =&-\frac{1}{4} \left( G^a_{\mu\nu}G^{a\mu\nu}\right) - \frac{1}{4}\lra{W_{\mu\nu} W^{\mu\nu}} - \frac{1}{4}B_{\mu\nu}B^{\mu\nu} -\frac{g_s^2}{16\pi^2}\theta_s \left( G^a_{\mu\nu}\tilde{G}^{a\mu\nu} \right) \notag \\
& + \frac{1}{2}\partial_\mu h \partial^\mu h -V(h) -\frac{v^2}{4}\lra{\bfv_\mu \bfv^\mu}\mathcal{F}_C(h) - \frac{v^{2}}{4} \lra{\bft \bfv_{\mu}}\lra{\bft \bfv^{\mu}} \mathcal{F}_{T}(h) \notag \\
& +i\bar{Q}_L\slashed{D} Q_L + i\bar{Q}_R\slashed{D}Q_R + i\bar{L}_L\slashed{D}L_L + i\bar{L}_R \slashed{D}L_R \notag \\
& -\frac{v}{\sqrt{2}}(\bar{Q}_L \bfu \mathcal{Y}_R^Q(h) Q_R) + h.c.) -\frac{v}{\sqrt{2}}(\bar{L}_L \bfu \mathcal{Y}_R^L(h)L_R + h.c.)\,,
\end{align}
where the subscript 2 indicates the chiral dimension of leading Lagrangian is 2, which will be discussed in the next subsection. Terms in the first line are the dynamic terms of gauge bosons and the theta term, and the second line contains the dynamic terms of NGBs, physical Higgs $h$ and its potential. The third line and the forth line describes the dynamic term and the mass terms of the fermions. The Yukawa coupling matrix $\mathcal{Y}^{Q/L}_R$ takes the form in the Eq.~\ref{eq:yukawa}. $\mathcal{F}_C(h)$ and $\mathcal{F}_T(h)$ appearing in the second lines are dimensionless polynomials of Higgs $h$.

\subsection{Power Counting and Higher Order Lagrangian}
\label{ssec:powe}

The power-counting of the HEFT is similar to the one in the chiral perturbation theory (ChPT) using the chiral dimension $d_\chi$~\cite{Weinberg:1978kz,Gasser:1983yg,Gasser:1984gg}, with certain improvements~\cite{Buchalla:2016sop,Gavela:2016bzc,Hirn:2005fr,Buchalla:2013eza,Buchalla:2012qq,Krause:2018cwe,Pich:2015kwa,Pich:2016lew}. Setting the LO Lagrangian be of the chiral dimension 2, the chiral dimensions of all the building blocks should be determined as follows. 
\begin{itemize}
    \item The gauge bosons $X_{\mu\nu}=G_{\mu\nu},W_{\mu\nu},B_{\mu\nu}$ are of $d_\chi=1$, and the derivatives $D$ or $\partial$ are of $d_\chi=1$. 
    \item The chiral dimension of the gauge coupling constants is 1, thus the dimension of the gauge vector fields are actually zero.
    \item The chiral dimension of $\bfv_\mu$ is 1, while the NGBs matrix $\bfu$ carries no chiral dimension.
    \item Every fermion doublet is of the chiral dimension 1/2, and the Yukawa coupling constants carry the chiral dimension 1.
    \item In particular, we make the convention that the spurion $\bft$ is of no chiral dimension, since this would describe the possible non-decoupling effects at the LO, for example, the triplet Higgs could develop a not-so-small VEV and causes custodial symmetry breaking effects at the LO Lagrangian, while in some literature such as Ref~\cite{Pich:2016lew,Krause:2018cwe,Pich:2018ltt}, the spurion is taken to be dimensional, in which the custodial symmetry breaking effects are always taking to be small.
\end{itemize}
This power-counting scheme is also consistent with the loop expansion~\cite{Buchalla:2013eza,Gavela:2016bzc,Pich:2016lew,Krause:2018cwe,Buchalla:2012qq,Hirn:2005fr}. 
Based on the discussion above, a general type of operators in the HEFT can be denoted by
\bea
{\kappa}^{k_i} \psi^{F_i} X^{V_i}_{\mu\nu} \bfu h D^{d_i}\bft^{S_i}\,,
\eea
where the number $k_i$ of the gauge or Yukawa couplings $\kappa$, $F_i$ the fermion fields $\psi$, $V_i$ the field-strength tensor $X_{\mu\nu}$, $d_i$ the covariant derivatives $D$, $S_i$ the spurions, and an arbitrary number of both the NGBs $\bfu$ and the Higgs boson $h$. 
The total chiral dimension of such type of operators reads
\bea
d_\chi = d_i + k_i + \frac{F_i}{2} + V_i \, (+ S_i)= 2 L_i + 2
\eea
where the spurion $\bft$ is taken to be dimensionless (dimensional). Furthermore, assuming the order of the chiral expansion is the same order as the one of the loop expansion, the total chiral dimension also determines the order of the loop expansion $L_i$. 

The above power counting on the gauge and Yukawa couplings implies that 
\bea 
\frac{p^{2}}{16 \pi^{2} v^{2}} \sim \frac{g^2}{(4 \pi)^{2}}, \,\frac{y^2}{(4 \pi)^{2}}, \, \frac{\lambda}{(4 \pi)^{2}} \ll 1\,.
\eea
Similar argument applies on the cases where the fermions are weakly coupled. Thus for higher-order operators, every fermion bilinear $(\overline{\psi}\psi)$ and gauge field strength tenser $X_{\mu\nu}\,,X=G,W,B$ carries chiral dimension 2 because of the weak coupling constants, for example, both the class $\psi^6Uh$ and $\psi^4XUh$ are of chiral dimension 6.
In the constructions of higher-dimension operators, we absorb the gauge and fermion coupling constants into the redefintion of the gauge field strength tensor $X$ and the fermions $\psi$, and thus redefine the chiral power of $X$ and $\psi$ as $2$ and $1$, respectively. Thus, the coupling constants in the higher-dimension operators are always implicit.  

We summarise the chiral dimensions of the building blocks in Tab~\ref{tab:buildingblocks}, and the operator classes up to $d_\chi=6$ are listed in Tab~\ref{tab:types}. We identify the operators of $d_\chi = 3,4$ as the NLO operators and those of $d_\chi = 5,6$ as the NNLO operators. In particular, the triple-gauge-boson type $X^3$ is excluded from the NNLO classes in this paper and is considered as NLO~\cite{Sun:2022ssa}.
The classes listed in this paper respect the convention that each of them contains a factor $Uh$ since $\bfu(x)$ is used to redefine the building blocks and $h(x)$ can be used freely in the operator constructions, as explained in Section~\ref{sec:next}. Because the spurion $\bft$ is of no chiral dimension, thus we do not write them in the classes explicitly, while it should be understood that each class in Tab~\ref{tab:types} contains all possible sub-classes with different numbers of spurions.

\begin{table}
    \centering
    \begin{tabular}{|c|c|c|}
    \hline
    $d_\chi$ & fermion sector & boson sector \\
    \hline
    3 & $\psi^2 UhD$ & \\
    \hline
    4 & $\psi^2XUh,\psi^4Uh,\psi^2UhD^2$ & $X^2Uh,XUhD^2,UhD^4$\\
    \hline
    5 & $\psi^2XUhD,\psi^4UhD,\psi^2UhD^3$ & \\
    \hline
    6 & $\psi^2X^2Uh,\psi^4XUh,\psi^6Uh,\psi^2XUhD^2,\psi^4UhD^2,\psi^2UhD^4$ & $X^3Uh,X^2UhD^2,XUhD^4,UhD^6$ \\
    \hline
    \end{tabular}
    \caption{Operator types of HEFT up to $d_\chi=6$. The classes listed in this paper respect the convention that each of them contains a factor $Uh$ since $\bfu(x)$ is used to 'dress' building blocks and $h(x)$ can be used freely in the operator constructions, as explained in Section~\ref{sec:next}. Because the spurion $\bft$ is of no chiral dimension, thus we do not write them in the classes explicitly, while it should be understood that each class in Tab~\ref{tab:types} contains all possible sub-classes with different numbers of spurions.}
    \label{tab:types}
\end{table}


There have been many discussions about the Lagrangian of the HEFT since the last century~\cite{Appelquist:1980vg,Longhitano:1980tm,Longhitano:1980iz,Dobado:1989ax,Feruglio:1992wf,Herrero:1993nc,Herrero:1994iu,Grinstein:2007iv}. Recently, the NLO operators have been constructed~\cite{Alonso:2012px,Buchalla:2012qq,Brivio:2013pma,Buchalla:2013eza,Buchalla:2013rka,Gavela:2014vra,Pich:2015kwa,Brivio:2016fzo,Merlo:2016prs,Pich:2016lew,Krause:2018cwe,Pich:2018ltt}, but none of them presents the complete and independent operator set, and the full flavor structures have never been considered. In Ref.~\cite{Sun:2022ssa}, the complete result of NLO operators are presented by the Young tensor method~\cite{Li:2020xlh,Li:2020gnx,Li:2022tec}, which is also used in this paper. At the NLO, there are 237 (8595) operators for one (three) generation fermions without right-handed neutrino, and 295 (11307) operators for one (three) generation fermions with right-handed neutrino. In this work, we construct the complete and independent NNLO operators for the first time. There are 12 classes in this order, ranging from chiral dimension 5 to 6, and the numbers of operators in each class are listed in the Tab~\ref{tab:nnlonumb}, and the total number of operators are $\frac{1}{9}(3672+25547{n_f}^2+420{n_f}^3+56684{n_f}^4+102{n_f}^5+17129{n_f}^6)$, corresponding to 11506 (1927574) for one (three) generations of fermions. In section \ref{sec:next} all these operators will be presented. 


\section{The Strategy of the Basis Construction}
\label{sec:onsh}

The following difficulties are present for the task of enumerating the NNLO operator basis:
\bit
\item The usual redundancy relations for operators, such as the Equation of Motion (EOM), Integration by Part (IBP), the Covariant Derivative Commutator (CDC) and various operator identities like the Fierz rearrangement and the Cayley-Hamilton relation.
\item The non-linear symmetry for the NGB impose constraints on the operators.
\item The operators in the broken phase organized in terms of spurions need special care regarding the group structures.
\eit

To tackle the first one, we briefly summarize the Young Tensor technique in sec.~\ref{sec:ampl}, which was implemented by Mathematica package and applied to various EFT's~\cite{Li:2022tec}. The non-linear symmetry of the NGB is taken care of by imposing the Adler zero conditions on the corresponding amplitudes, as explained in sec.~\ref{ssec:adle}. Finally in sec.~\ref{sec:spurion} we elaborate the treatment for the spurions in order to systematically organize the operators in the symmetry broken phase.


\begin{table}
\begin{center}
\begin{tabular}{|c|c|c|c|c|c|}
\hline
building blocks & spinor formalism & Lorentz group & $SU(2)_L$ & $SU(3)_C$ & $d_\chi$ \\
\hline
$L_L$ & ${L_L}_{\alpha}$ & $(\frac{1}{2},0)$ & Fundamental & Singlet & 1 \\
\hline
$L_R$ & ${L_R}^{\dot{\alpha}}$ & $(0,\frac{1}{2})$ & Fundamental & Singlet & 1 \\
\hline
$Q_L$ & ${Q_L}_{\alpha}$ & $(\frac{1}{2},0)$ & Fundamental & Fundamental & 1 \\
\hline
$Q_R$ & ${Q_R}^{\dot{\alpha}}$ & $(0,\frac{1}{2})$ & Fundamental & Fundamental & 1 \\
\hline
$W_L$ & ${W_L}_{\alpha\beta}$ & $(1,0)$ & Adjoint & Singlet & 2 \\
\hline
$W_R$ & ${W_R}^{\dot{\alpha}\dot{\beta}}$ & $(0,1)$ & Adjoint & Singlet & 2 \\
\hline
$G_L$ & ${G_L}_{\alpha\beta}$ & $(1,0)$ & Singlet & Adjoint & 2 \\
\hline
$G_R$ & ${G_R}^{\dot{\alpha}\dot{\beta}}$ & $(0,1)$ & Singlet & Adjoint & 2 \\
\hline
$B_L$ & ${B_L}_{\alpha\beta}$ & $(1,0)$ & Singlet & Singlet & 2 \\
\hline
$B_R$ & ${B_R}^{\dot{\alpha}\dot{\beta}}$ & $(0,1)$ & Singlet & Singlet & 2 \\
\hline
$\bfv^\mu\sim D^\mu\phi$ & ${(D\phi)}_{\dot{\alpha}\beta}$ & $(\frac{1}{2},\frac{1}{2})$ & Adjoint & Singlet & 1 \\
\hline
$D^\mu$ & $D_{\alpha\dot{\beta}}$ & $(\frac{1}{2},\frac{1}{2})$ & Singlet & Singlet & 1 \\
\hline
$\bft$ & $\bft$ & $(0,0)$ & Adjoint & Singlet & 0 \\
\hline
\end{tabular}
\end{center}
\caption{The building blocks of HEFT, their representation under the Lorentz and gauge groups, and the chiral dimension of them. To satisfy the Adler zero condition, we use $D^\mu\phi$ replacing $\bfv^\mu$. }
\label{tab:buildingblocks}
\end{table}

\subsection{Review on Young Tensor Method}
\label{sec:ampl}
An EFT operator should be singlet under both the Lorentz and gauge groups. For a specific field content, the independent Lorentz and gauge structures are of finite dimension, thus span two independent linear spaces respectively, the Lorentz space and the gauge space, in which the independent structures are called the Lorentz basis and the gauge basis. The whole space spanned by the independent operators constructed from this field content are the tensor product of these two spaces. 

In the Young tensor method, the Lorentz basis of operators are related to the corresponding basis of local on-shell amplitudes. 
The on-shell solutions of the fields $\phi_i$ are given by the spinor-helicity variables $(\lambda_{i\alpha},\tilde\lambda_i^{\dot\alpha})$ according to their helicities $h_i$, and thus we obtain the correspondence
\be
(D^{r_i-h_i})\phi_i \sim \lambda_i^{r_i-h_i}\tilde\lambda^{r_i+h_i}
\ee
with free spinor indices. Due to the Lorentz invariance, these indices are contracted by invariant tensors $\epsilon_{\alpha\beta}$ and $\tilde\epsilon_{\dot\alpha\dot\beta}$ under the $SU(2)_l$ and $SU(2)_r$ subgroups of the Lorentz symmetry, with the numbers
\be
n = \frac12\sum_i(r_i-h_i) \equiv \frac{r-h}{2}\ ,\quad \tilde{n} = \frac12\sum_i(r_i+h_i) \equiv \frac{r+h}{2}\ .
\ee
where $r=\sum_i r_i$ and $h=\sum_i h_i$ are defined. The contracted spinor variables are denoted as usual
\be
\lambda_i^\alpha\lambda_{j\alpha} = \langle ij\rangle,\quad \tilde{\lambda}_{i\dot{\alpha}}\tilde{\lambda}_j^{\dot{\alpha}} = [ij]\,.
\ee
Therefore, any operator could be mapped to a unique (combination of) on-shell amplitude in terms of the spinor variables which can be written as
\be
\mathcal{O} = \bigotimes_{n}\epsilon\bigotimes_{\tilde{n}}\tilde{\epsilon}\prod_i (D^{r_i-h_i})\phi_i \quad\sim\quad \mathcal{M} = \langle\cdot\rangle^{\otimes n}[\cdot]^{\otimes \tilde{n}}
\ee

\comments{
The Lorentz group can be complexified as product of two $SU(2)$ groups
\be
SO(1,3) = SU(2)_l\times SU(2)_r\,,
\ee
then all the representations of the Lorentz group can be labeled by the casimirs of $SU(2)_l$ and $SU(2)_r$, $(j_l,j_r)$. We follow the convention that the fundamental indices of $SU(2)_l$ are the undotted Greek letters and those of $SU(2)_r$ are the dotted Greek letters. The fundamental representations of these $SU(2)_l$ and $SU(2)_r$ groups are usually called left- and right-handed spinors. In the Table~\ref{tab:buildingblocks} we list all the building blocks' representations in the HEFT. 

Considering that the fundamental building blocks are fields with one or more derivatives applied, $(D^{r_i-h_i})\phi_i$, where $r_i$ is the half number of the spinor indices of this building block, and $h_i$ is the helicity of the field $\phi_i$, and all operators are the Lorentz scalar, which means all the Lorentz indices should be contracted by the spinor metric $\epsilon$ or $\tilde{\epsilon}$, an operator composed of these building blocks can be expressed as
\be
\label{op}
\mathcal{O} = \mathcal{T}\bigotimes_{n}\epsilon\bigotimes_{\tilde{n}}\tilde{\epsilon}\prod_i (D^{r_i-h_i})\phi_i\,,
\ee
where $\mathcal{T}$ is the group factor to form the invariant under the gauge group. The number of the tensor $\epsilon$, $n$, and the number of the tensor $\tilde{\epsilon}$, $\tilde{n}$ are determined by the field content,
\be
n=\frac{r-h}{2},\quad \tilde{n}=\frac{r+h}{2}\,,
\ee
where $r=\sum_i r_i$ and $h=\sum_i h_i$.

On the other hand, an amplitude $\mathcal{M}$ are composed of the left- and right-handed spinors $\lambda_i,\tilde{\lambda}_i$ as well as $\epsilon,\tilde{\epsilon}$ to contract all the indices,
\be
\lambda_i^\alpha\lambda_{j\alpha} = \langle ij\rangle,\quad \tilde{\lambda}_{i\dot{\alpha}}\tilde{\lambda}_j^{\dot{\alpha}} = [ij]\,,
\ee
and it should follow the little group transformation 
$\mathcal{M}\rightarrow e^{ih_i\varphi}\mathcal{M}$ for the $i$th particle, where $h_i$ is the helicity of this particles and $\varphi$ is a real phase. Thus such a massless particle of helicity $h_i$ contributes a factor $\lambda_i^{r_i-h_i}\tilde{\lambda}_i^{r_i+h_i}$ to the amplitude.
So a general amplitude takes the form that
\be
\mathcal{M} = \bigotimes_{n}\epsilon\bigotimes_{\tilde{n}}\tilde{\epsilon}\prod_i \lambda_i^{r_i-h_i}\tilde{\lambda}_i^{r_i+h_i}\,.
\ee
Comparing with Eq.~\ref{op}, we can make the identification that
\be
\mathcal{O}\rightarrow \mathcal{T} \mathcal{M} = \mathcal{T}\bigotimes_{n}\epsilon\bigotimes_{\tilde{n}}\tilde{\epsilon}\prod_i \lambda_i^{r_i-h_i}\tilde{\lambda}_i^{r_i+h_i}\,.
\ee
This identification induces the operator-amplitude correspondence. 



}

With the amplitude-operator correspondence, we start to construct the Lorentz basis.
For now, we consider the flavor-blind operators. At this stage, all the fields involved in the operators are taken to be different, and the basis obtained here are the so-called Lorentz y-basis, constructed by semi-standard Young tableaux (SSYT). The SSYTs are obtained by filling the primary Young diagram, which is completely determined by field contents. For example, the primary Young diagram of the Lorentz space takes the form
\begin{eqnarray}\label{eq:YD_shape}
\arraycolsep=0pt\def\arraystretch{1}
\rotatebox[]{90}{\text{$N-2$}}
	\left\{
	\begin{array}{cccccc}
		\yng(1,1) &\ \ldots{}&\ \yng(1,1)& \overmat{n}{\yng(1,1)&\ \ldots{}\  &\yng(1,1)} \\
		\vdotswithin{}& & \vdotswithin{}&&&\\
		\undermat{\tilde{n}}{\yng(1,1)\ &\ldots{}&\ \yng(1,1)} &&&
	\end{array}
	\right.\,,
	\\
	\nonumber 
\end{eqnarray}
where $N$ is the number of particles involved in the operator. The numbers of indices to fill in the primary Young diagram are determined by
\beq
\#i = \frac{1}{2}n_D+\sum_{h_i>0}|h_i|-2h_i,\quad i=1,2,\dots N\,,\label{eq:numi}
\eeq
where $n_D$ is the number of derivatives. The primary Young diagram corresponds to the singlet representation of the Lorentz group, while the non-singlet representation of the $SU(N)$ group. It has been proved that it is the primary Young diagram that eliminate the IBP redundancies~\cite{Li:2020xlh,Li:2020gnx,Li:2022tec}, which form the independent y-basis of the amplitudes/operators. The explicit forms of the basis amplitudes $\mathcal{B}_i$ can be directly obtained by translating the Semi-Standard Young Tableau (SSYT) of the primary Young Diagram to the spinor brackets. Besides the simplicity of the construction, the y-basis is also convenient for the decomposition of any given local amplitudes/operators, which we call the reduction to the y-basis
\be 
\mathcal{M} = \sum_i c_i \mathcal{B}_i.
\ee
It turns out to be crucial in the various manipulation of amplitudes and operators in the related studies.



\subsection{Adler Zero Condition on Amplitude Basis}
\label{ssec:adle}

Before the general effective Lagrangian of the nonlinearly realised symmetry developed by CCWZ~\cite{Coleman:1969sm,Callan:1969sn}, Adler~\cite{Adler:1964um,Adler:1965ga} derived that the scattering amplitude with single pion emission vanishes in the soft limit of the pion momentum, which is called the Adler zero condition. With the CCWZ formalism, the pseudoscalar pions are considered as the NGBs after symmetry breaking, and the Adler zero condition is trivially fulfilled in the effective Lagrangian in the CCWZ formalism. But the Young tensor method starts from the general Lorentz structure, instead of the pion matrix field in the CCWZ formalism. Therefore, not all the Lorentz structures in the Young tensor satisfy Adler zero condition. According to the amplitude-operator correspondence, the Lorentz structures satisfying the Adler zero condition corresponds to the amplitudes satisfying soft limit \cite{Low:2014nga,Low:2014oga,Cheung:2014dqa,Cheung:2015ota,Low:2019ynd,Dai:2020cpk}.


In practice, we adopt the scalar field $\phi$ which is the adjoint representation under the $SU(2)_L$, and since it is the NGBs, the Adler zero condition implies that there is at least 1 derivative applied on $\phi$. This is always possible by using the IBP relation  (for amplitude, it is the momentum conservation relation $\sum_i\langle li\rangle[ik] = 0$). Thus in this method, we use $D^\mu\phi$ replacing $\bfv^\mu$, which can be regarded as the leading term of $\bfv^\mu$ defined in Eq.~\ref{eq:vmu}. In the following, we will present the procedure of imposing the Adler zero condition which is also shown in Ref.~\cite{Low:2022iim, Sun:2022ssa}.

Let us consider a type of operators with $N$ particles, including at least one NGB $\pi$. Based on the Young tensor method above, the Lorentz basis can be expressed as the $N$-point on-shell amplitudes  $\{\calb^{(N)}_i,i=1,2,\dots d_N\}$, where $d_N$ is the dimension of this Lorentz basis. In terms of such basis, any Lorentz structure of this type takes the form that
\beq
\calm^{(N)} = \sum_{i=1}^{d_N} c_i\calb^{(N)}_i, 
\eeq
where $c_i$ are coefficients under this basis. If this amplitude satisfies the Adler zero condition, it vanishes when the external pion momentum $p_\pi$ becomes soft:
\beq
\calm^{(N)}( p_\pi \rightarrow 0) = 0 =\sum_{i=1}^{d_N}c_i\calb^{(N)}_i( p_\pi \rightarrow 0).
\label{eq:soft}
\eeq
Here $\calb^{(N)}_i( p_\pi \rightarrow 0)$ becomes $(N-1)$-point on-shell amplitudes, which are generally not independent and can be expanded by the $(N-1)$-point basis with the soft particle $\pi$ removed, $\{\mathcal{B}^{(N-1)}_i(\bar{\pi}),i=1,2,\dots d_{N-1}\}$, where $d_{N-1}$ is the dimension of such $(N-1)$-point Lorentz basis, and $\bar{\pi}$ implies that the soft particle $\pi$ is removed,
\begin{equation}
\calb^{(N)}_i( p_\pi \rightarrow 0) = \sum_{j=1}^{d_{N-1}}f_{ij}\mathcal{B}^{(N-1)}_i(\bar{\pi})\,.
\end{equation}

Furthermore, since the removed $\pi$ is a scalar particles, all the $(N-1)$-point basis $\mathcal{B}^{(N-1)}_i(\bar{\pi})$ can be expanded by the original $N$-point basis $\calb^{(N)}_i$
\begin{equation}
    \mathcal{B}^{(N-1)}_i(\bar{\pi}) = \sum_{l=1}^{d_N}d_{il}\calb^{(N)}_i\,.
\end{equation}
Combining several equations above, we can obtain the expansion
\beq
0=\sum_{l=1}^{d_N}(\sum_{l=i}^{d_N}c_i\mathcal{{K}}_{il})\calb^{(N)}_l\,,
\eeq
where $\mathcal{K}_{il}$ is the expansion matrix
\be
\mathcal{K}_{il} = \sum_{j=1}^{d_{N-1}}f_{ij}d_{jl}\,.\label{eq:kil}
\ee
Since the basis $\{\calb^{(N)}_i,i=1,2,\dots d_N\}$ are independent, this equation holds only if all the coefficients vanish,
\beq
0=\sum_{i=1}^{d_N}c_i\mathcal{K}_{ij},\quad (j=1,2,\dots d_N)\,. \label{eq:sysoflin}
\eeq
This is a system of linear equations about $c_i$, whose solutions span the subspace satisfying the Adler zero condition, which constitute the amplitude basis involving NGBs.  If there are more that one NGBs, there is a system of linear equations for each of them, and the structures satisfying the Adler zero condition are their common solutions.

Let us present an explicit example of a five-particle class ${F_L}\phi^4D^4$, with the helicities that $\{-1,0,0,0,0\}$. We suppose that there are 4 derivatives in this class, and the second to the forth spinless particles are NGBs. According to the Young tensor method, we can get the complete Lorentz basis is of 14 dimension,
\begin{align}
    \calb_1 &= -\lrs{45}^2\lra{45}\lra{14}\lra{15}\,,\quad \calb_2 = -\lrs{34}\lrs{35}\lra{13}^2\lra{45}\,,\notag \\
    \calb_3 &= \lrs{35}\lrs{45}\lra{45}\lra{13}\lra{14}\,,\quad \calb_4 = -\lrs{34}\lrs{45}\lra{45}\lra{13}\lra{14}\,,\notag \\
    \calb_5 &= \lrs{34}\lrs{45}\lra{14}^2\lra{35}\,,\quad \calb_6 = \lrs{45}\lrs{35}\lra{35}\lra{14}\lra{15}\,,\notag \\
    \calb_7 &= -\lrs{35}^2\lra{35}\lra{13}\lra{15}\,,\quad \calb_8 = -\lrs{34}^2\lra{34}\lra{13}\lra{14}\,,\notag \\
    \calb_9 &= \lrs{34}\lrs{35}\lra{35}\lra{13}\lra{14}\,,\quad \calb_{10} = \lrs{35}\lrs{25}\lra{25}\lra{13}\lra{15}\,,\notag \\
    \calb_{11} &= \lrs{34}\lrs{24}\lra{24}\lra{13}\lra{14}\,,\quad \calb_{12} = \lrs{24}\lrs{35}\lra{13}\lra{14}\lra{25}\,,\notag \\
    \calb_{13} &= -\lrs{24}\lrs{45}\lra{14}^2\lra{25}\,,\quad \calb_{14} = -\lrs{45}\lrs{25}\lra{25}\lra{14}\lra{15}\,,
\end{align}
but not all of them satisfy the Adler zero condition. Since there are 3 NGB particles in this class, their soft limits should be taken separately. For the second particle, we take the momentum $p_2\rightarrow 0$, which is equivalent to the condition $|2\rangle,|2] \rightarrow 0$, the amplitude after the limitation
\begin{equation}
\label{eq:b1oft}
    \mathcal{B}_1 \rightarrow  \mathcal{B}_1  (p_2 \to 0) = -\lrs{34}^2\lra{34}\lra{13}\lra{14}
\end{equation}
should belong to the type $\{-1,0,0,0\}$.
Note that the $4$-point Lorentz basis of the type $\{-1,0,0,0\}$ contains only 2 Lorentz structures
\begin{equation}
    B_1(\bar{2}) = \lra{13}\lra{14}\lra{34}\lrs{34}^2\,,\quad B_2(\bar{2}) = -\lra{13}\lra{14}\lra{24}\lrs{24}\lrs{34}\,.
\end{equation}
Given that the two basis can be expanded in the original 5-point basis $d_{1i} = \delta_{i1}, d_{2i} = \delta_{i6}$, where $i = 1, \cdots, 14$ and 
the soft amplitude $\mathcal{B}_1 (p_2 \to 0)$ can be expanded in the two 4-point basis $f_{11} = 1, f_{12} = 0$
the soft amplitude can be expanded in the original 5-point basis as
\begin{align}
    \mathcal{B}_1 &\rightarrow -\lrs{34}^2\lra{34}\lra{13}\lra{14} = \sum_{i=1}^{14}\sum_{j=1}^2f_{1j}d_{ji}\mathcal{B}_i \notag \\
    &= \sum_{i=1}^{14}f_{11}d_{1i}\mathcal{B}_i = \sum_{i=1}^{14}1\times \delta_{i1}\mathcal{B}_i  = \mathcal{B}_1\,,
\end{align}
Similar procedure can be applied on the other 5-point basis, and finally, we can get the full matrix $\mathcal{K}$ that
\begin{equation}
    \mathcal{K}= \left(
\begin{array}{cccccccccccccc}
 1 & 0 & 0 & 0 & 0 & 0 & 0 & 0 & 0 & 0 & 0 & 0 & 0 & 0 \\
 1 & 0 & 0 & 0 & 0 & 0 & 0 & 0 & 0 & 0 & 0 & 0 & 0 & 0 \\
 1 & 0 & 0 & 0 & 0 & 0 & 0 & 0 & 0 & 0 & 0 & 0 & 0 & 0 \\
 1 & 0 & 0 & 0 & 0 & 0 & 0 & 0 & 0 & 0 & 0 & 0 & 0 & 0 \\
 0 & 0 & 0 & 0 & 0 & 1 & 0 & 0 & 0 & 0 & 0 & 0 & 0 & 0 \\
 0 & 0 & 0 & 0 & 0 & 1 & 0 & 0 & 0 & 0 & 0 & 0 & 0 & 0 \\
 0 & 0 & 0 & 0 & 0 & 1 & 0 & 0 & 0 & 0 & 0 & 0 & 0 & 0 \\
 -1 & 0 & 0 & 0 & 0 & 1 & 0 & 0 & 0 & 0 & 0 & 0 & 0 & 0 \\
 0 & 0 & 0 & 0 & 0 & 1 & 0 & 0 & 0 & 0 & 0 & 0 & 0 & 0 \\
 0 & 0 & 0 & 0 & 0 & 0 & 0 & 0 & 0 & 0 & 0 & 0 & 0 & 0 \\
 0 & 0 & 0 & 0 & 0 & 0 & 0 & 0 & 0 & 0 & 0 & 0 & 0 & 0 \\
 0 & 0 & 0 & 0 & 0 & 0 & 0 & 0 & 0 & 0 & 0 & 0 & 0 & 0 \\
 0 & 0 & 0 & 0 & 0 & 0 & 0 & 0 & 0 & 0 & 0 & 0 & 0 & 0 \\
 0 & 0 & 0 & 0 & 0 & 0 & 0 & 0 & 0 & 0 & 0 & 0 & 0 & 0 \\
\end{array}
\right)\,,
\end{equation}
and thus the non-trivial equations in Eq.~\ref{eq:sysoflin} are
\begin{equation}
\left\{
    \begin{array}{l}
    c_1 + c_2 + c_3 + c_4 - c_8 = 0\\
    c_5 + c_6 + c_7 + c_8 + c_9 = 0
    \end{array}
\right.
\end{equation}
Similarly, we can get the equations for the third and forth soft particles,
\begin{equation}
\left\{
    \begin{array}{l}
    c_1  = 0\\
    c_{13}+c_{14} = 0
    \end{array}
\right.
\,,\quad \left\{
    \begin{array}{l}
    c_7 = 0 \\
    c_{10} = 0
    \end{array}
\right.\,.
\end{equation}
There are totally 6 constraints, and the solution space is of dimension 8. Thus there are 8 independent Lorentz basis satisfying the Adler zero condition, we present the transforming matrix that
\begin{equation}
    \left(
\begin{array}{cccccccccccccc}
 0 & 0 & 0 & 0 & 0 & 0 & 0 & 0 & 0 & 0 & 0 & 0 & -1 & 1 \\
 0 & 0 & 0 & 0 & 0 & 0 & 0 & 0 & 0 & 0 & 0 & 1 & 0 & 0 \\
 0 & 0 & 0 & 0 & 0 & 0 & 0 & 0 & 0 & 0 & 1 & 0 & 0 & 0 \\
 0 & 0 & 0 & 0 & -1 & 0 & 0 & 0 & 1 & 0 & 0 & 0 & 0 & 0 \\
 0 & 1 & 0 & 0 & -1 & 0 & 0 & 1 & 0 & 0 & 0 & 0 & 0 & 0 \\
 0 & 0 & 0 & 0 & -1 & 1 & 0 & 0 & 0 & 0 & 0 & 0 & 0 & 0 \\
 0 & -1 & 0 & 1 & 0 & 0 & 0 & 0 & 0 & 0 & 0 & 0 & 0 & 0 \\
 0 & -1 & 1 & 0 & 0 & 0 & 0 & 0 & 0 & 0 & 0 & 0 & 0 & 0 \\
\end{array}
\right)\,.
\end{equation}
Therefore, the resulting basis are usually polynomials  of the original SSYT basis,
\begin{align}
    \calb'_1 &= \lrs{24}\lrs{45}\lra{14}^2\lra{25} - \lrs{45}\lrs{25}\lra{25}\lra{14}\lra{15}\,,\notag \\
    \calb'_2 &= \lrs{24}\lrs{35}\lra{13}\lra{14}\lra{25}\,,\notag \\
    \calb'_3 &= \lrs{34}\lrs{24}\lra{24}\lra{13}\lra{14}\,,\notag \\
    \calb'_4 &= -\lrs{34}\lrs{45}\lra{14}^2\lra{35} + \lrs{34}\lrs{35}\lra{35}\lra{13}\lra{14}\,,\notag \\
    \calb'_5 &= -\lrs{34}\lrs{45}\lra{14}^2\lra{35} + \lrs{34}^2\lra{34}\lra{13}\lra{14}\,,\notag \\
    \calb'_6 &= -\lrs{34}\lrs{45}\lra{14}^2\lra{35} + \lrs{45}\lrs{35}\lra{35}\lra{14}\lra{15}\,,\notag \\
    \calb'_7 &= \lrs{34}\lrs{35}\lra{13}^2\lra{45} -\lrs{34}\lrs{45}\lra{45}\lra{13}\lra{14}\,,\notag \\
    \calb'_8 &=  \lrs{34}\lrs{35}\lra{13}^2\lra{45} + \lrs{35}\lrs{45}\lra{45}\lra{13}\lra{14}\,.\label{eq:examp2}
\end{align}




\subsection{Spurion Technique for the Gauge Structure}
\label{sec:spurion}

It is known in the group theory that every $SU(N)$ irreducible representation (irrep) corresponds to a Young diagram too. For example, the typical irreps of the $SU(2)$ and $SU(3)$ groups are 
\begin{center}
\begin{tabular}{ll|ll}
\hline
$SU(2)$ & & SU(3) &   \\
\hline
& $t_i \in \mathbf{2} \sim \yng(1)$ & & $t_a \in \mathbf{3} \sim \yng(1)$ \\
& $\epsilon_{ij}t^j \in \overline{\mathbf{2}} \sim \yng(1)$ & & $\epsilon_{abc}t^c\in \overline{\mathbf{3}} \sim \yng(1,1)$ \\
& $t^I{\tau^{Ik}_i}\epsilon_{kj} \in \mathbf{3} \sim \yng(2)$ & & $t^A{\lambda^{Ad}}_a\epsilon_{dbc} \in \mathbf{8} \sim \yng(2,1)$ \\
\hline
\end{tabular}
\end{center}
where $\tau,\lambda$ are the generators of the $SU(2), SU(3)$ respectively. 
In the EFTs with unbroken symmetries, like the SMEFT, the effective operators belong to the singlets under the gauge groups $SU(N)$, represented by Young diagrams with only $N$-row columns,
\begin{equation}
    SU(2)\sim \yng(2,2)\dots\yng(1,1)\,,\quad SU(3)\sim \yng(2,2,2)\dots\yng(1,1,1)\,.
\end{equation}
In the Young tensor method, we construct the independent basis of gauge structures by decomposing the tensor product of the Young diagrams of the dynamical fields following the Littlewood-Richardson rule (LR rule)~\cite{Li:2020xlh,Li:2020gnx,Li:2022tec}, and select all the singlet terms therein. 

\comments{
For example, considering a type of operators in which there are 4 quarks $Q^2{Q^\dagger}^2$, their $SU(3)$ representations are
\begin{align}
Q_a \sim \young(a)\,, \quad 
\epsilon_{bca}{Q^\dagger}^a \sim \young(b,c)\,,
\end{align}
and the outer product of them are
\begin{equation}
    \young(\aone)\,\xrightarrow{\young(\atwo)}\,\left\{\begin{array}{l}
    \young(\aone\atwo)\,\xrightarrow{\young(\bthree,\cthree)}\,\young(\aone\atwo,\bthree,\cthree)\,\xrightarrow{\young(\bfour,\cfour)}\,\young(\aone\atwo,\bthree\bfour,\cthree\cfour)\,\rightarrow\, \epsilon^{a_1b_2c_3}\epsilon^{a_2b_4c_4}  \\
    \\
    \young(\aone,\atwo)\,\xrightarrow{\young(\bthree,\cthree)}\,\left\{\begin{array}{l}
    \young(\aone\bthree,\atwo,\cthree)\, \xrightarrow{\young(\bfour,\cfour)}\,\young(\aone\bthree,\atwo\bfour,\cthree\cfour)\,\rightarrow\, \epsilon^{a_1a_2c_3}\epsilon^{b_3b_4c_4} \\
    \\
    \young(\aone\bthree,\atwo\cthree)\,\xrightarrow{\young(\bfour,\cfour)}\, 0
    \end{array}\right.
    \end{array}\right.\,,
\end{equation}
where only the singlet Young diangrams are kept. Thus there are two independent $SU(3)$ bases for this type.
}



However, in the HEFT, we are dealing with effective operators in the broken phase of the custodial symmetry.
The breaking pattern is characterized by a spurion field $\bft$ under the adjoint representation of the $SU(2)_L$ group, which enters the gauge and Yukawa interactions.
Typically the spurion fields are introduced in theories with broken symmetries to parametrize the symmetry breaking effects. In the unbroken phase, the spurion is a dynamical degree of freedom, which makes sure the operators in the unbroken phase to be singlet under the correspondent symmetry. In the broken phase, the spurion degree of freedom is frozen, and thus the operators in the broken phase becomes non-singlet under the same symmetry. Spurions are thus the auxiliary fields that contract with the operators to make a singlet, which are supposed to take VEV after the symmetry breaking, thus are frozen degrees of freedom, and do not participate in the Lorentz structures, which being said, we should not treat them as scalar building blocks in the Lorentz sector because the derivatives can not act on them. In practice, for a type of operators involving the spurions, they should be deleted from the helicity list, and their generators of symmetric group are taken to be identity matrix.


Based on the above, the spurions are only used in constructing the group factor of the corresponding symmetry.
Multiple copies of spurions may be used to construct operators under different non-singlet representations, and in turn there may be different combinations of spurions that constitute the same representation. For example, suppose we have a spurion $\mathbf{T}^a$ under the adjoint representation $\mathbf{8}$ of $SU(3)$, hence we have
\begin{equation}
    \mathbf{T}^a\,\in\,\mathbf{8}\ ,\quad d^{abc}\mathbf{T}^b\mathbf{T}^c\,\in\,\mathbf{8}\ ,
\end{equation}
where $d^{abc}$ is the total symmetric tensor. Both the 2 combinations above are capable of contracting with an operator $\mathcal{O}^a$ under the adjoint representation of $SU(3)$. It is obvious that the first would be the dominant one, and the second is sub-leading, since it takes more spurions. Thus we do not count the second one for the fixing non-singlet operator $\calo^a$ in the broken phase, which is equivalent to the eliminations of the gauge structures like $d^{abc}\bft^b\bft^c$ in the type with 2 spurions.
Another restriction is the identity among the same kind of spurions, which demands totally symemtric combinations, thus $f^{abc}\mathbf{T}^b\mathbf{T}^c$, where $f^{abc}$ is the totally anti-symmetric tensor, is zero. 
Furthermore, for a general type with $j$ spurions, the only reserved structures are them of the traceless and totally symmetric combination,
\begin{equation}
\label{eq:spinj}
    \mathbf{T}^{\{I_1}\cdots\mathbf{T}^{I_j\}}\,\in\, \text{spin }j\,.
\end{equation}

Thus we need consider the representation of spurions and other fields separately, and both the Young diagrams of them are non-singlet, but their outer product can form singlet Young diagram. In particular, the SSYTs of the spurions representation should be symmetric under the permuations among the indices from different spurions. In general, this is difficult, but in the case of the $SU(2)$ group, it is quite straightforward to deal with, as shown in below.

The spurion $\bft^I$ in the HEFT belongs to the adjoint representation of the $SU(2)$ group,
\begin{equation}
    \bft^I{\tau^{Ik}}_i\epsilon_{kj} \in \young(ij)\,,
\end{equation}
where the two indices are symmetric. The tensor product of the 2 spurions can be expressed by the Young diagrams
\begin{equation}
    \yng(2) \otimes \yng(2) = \yng(3,1) \oplus \yng(2,2) \oplus \yng(4)\,,
\end{equation}
and only the last one is traceless and totally symmetric,
\begin{equation}
    \young(ijmn) = \bft^{\{I}\bft^{J\}}{\tau^I}^l_i{\tau^J}^k_m\epsilon_{lj}\epsilon_{kn}\,,
\end{equation}
since the first two diagrams contain columns of length 2, which correspond anti-symmetrical tensor $\epsilon$ contracting with the indices of spurions to generate traces, for example, the tensor in the first diagram 
\begin{equation}
    \young(ijm,n) = \bft^I\bft^J {\tau^I}^k_i{\tau^J}^l_m \epsilon_{kj}\epsilon_{ln}\epsilon^{in} = \bft^I\bft^J\epsilon^{IJK}{\tau^K}^k_m\epsilon_{kj}\,,
\end{equation}
which is actually zero since there is only single spurion field. The tensor of the second diagram takes the form  
\begin{equation}
    \young(ij,mn) =  \bft^I\bft^J {\tau^I}^k_i{\tau^J}^l_m \epsilon_{kj}\epsilon_{ln}\epsilon^{im}\epsilon^{jn} = \bft^I\bft^I\,,
\end{equation}
which is the self-contraction of spurions, and should be eliminated as well.
Furthermore, the tensor product of the 3 spurions takes the form 
\begin{equation}
    \yng(2) \otimes \yng(2) \otimes \yng(2) = \yng(3,3) \oplus \yng(5,1) \oplus 3\,\yng(4,2) \oplus \yng(6)\,,
\end{equation}
and only the last one is traceless and totally symmetric. Generally, the traceless and totally symmetric combinations of $j$ spurions make a spin-$j$ representation, which corresponds the irreducible representation with the highest weight in the direct-product decomposition of them, corresponding the diagram that
\begin{center}
\begin{tikzpicture}
\draw (0pt,0pt) rectangle (14pt,14pt);
\draw (14pt,0pt) rectangle (28pt,14pt);
\draw [loosely dotted] (30pt,7pt)--(40pt,7pt);
\draw (42pt,0pt) rectangle (56pt,14pt);
\draw (56pt,0pt) rectangle (70pt,14pt);
\draw [|<-] (0pt,21pt) -- (28pt,21pt);
\draw [|<-] (70pt,21pt) -- (42pt,21pt);
\node (2n) at (35pt,21pt) {$2j$};
\end{tikzpicture}
\end{center}
and the compensation formed by other dynamical fields of this representation to form the $SU(2)$ singlet takes the form 
   \begin{center}
    \begin{tikzpicture}
\draw (0pt,0pt) rectangle (14pt,14pt);
\draw (14pt,0pt) rectangle (28pt,14pt);
\draw [loosely dotted] (30pt,7pt)--(40pt,7pt);
\draw (42pt,0pt) rectangle (56pt,14pt);
\draw (0pt,14pt) rectangle (14pt,28pt);
\draw (14pt,14pt) rectangle (28pt,28pt);
\draw [loosely dotted] (30pt,21pt)--(40pt,21pt);
\draw (42pt,14pt) rectangle (56pt,28pt);
\draw (56pt,14pt) rectangle (70pt,28pt);
\draw (70pt,14pt) rectangle (84pt,28pt);
\draw [loosely dotted] (86pt,21pt)--(96pt,21pt);
\draw (98pt,14pt) rectangle (112pt,28pt);
\draw (112pt,14pt) rectangle (126pt,28pt);
\draw [|<-] (56pt,35pt) -- (84pt,35pt);
\draw [|<-] (126pt,35pt) -- (98pt,35pt);
\node (2j) at (91pt,35pt) {$2j$};
    \end{tikzpicture}
\end{center} 
which can be obtained by the application of the Littlewood-Richardson rule reversely. To eliminate the redundancies about the spurions mentioned above, we need do the outer product of the j copies of the spurions and the other dynamical fields separately to form two SSYTs of the shape above, then combine them back to form the singlet Young diagrams by simply binding the two SSYTs together.


\subsection{Flavor Structure involving Goldstone and Spurion}
\label{sec:pbas}


For both the Lorentz and gauge space, the y-basis are often polynomials and we can transform them to another basis in which all operators are monomials, the m-basis. The tensor product of the Lorentz and gauge m-basis constitutes the full space of all the independent flavor-blind operators, which is called the operator m-basis.

Considering the repeated fields in the operators, the operators in the m-basis obtained above are usually redundant. In this case, a field with flavor number $n_f$ can be regarded as a $n_f$-multiples of the flavor group $SU(n_f)$. If an operator has $n$ such fields, this operator behaves as a n-rank tensor under the group $SU(n_f)$. This flavor tensor-product space is fully divided into several disjoint subspaces, each of which furnishes an irreducible representation of the symmetric group $S_n$. The operators in every irreducible representation of the $S_n$ have specific permutation symmetries. Thus we introduce the p-basis composed by these operators. In the p-basis, not all the operators are physical, for example, if the repeated field has flavor number 1, all the operators in the p-basis are zero, except for those in the completely symmetric representation of $S_n$. Besides, if the p-basis contain operators with the mixed flavor symmetry such as ${\tiny\yng(2,1)}$, the irreducible subspace of $SU(n_f)$ marked by this Young diagram has multiplicity equal to the dimension of the irreducible representation of the symmetry group $S_{n}$ presented by the same Young diagram. It can be proved that these irreducible subspaces are isomorphic to each other \cite{Li:2022tec}, and only one of them needs to be reserved. After eliminating such redundancies, the remaining operators form the so-called f-basis or p'-basis, which serves as the final result.
In practice, p-basis can be obtained by applying the idempotent elements of group algebra $\tilde{S}_n$ on the m-basis.
\be
\mathcal{O}^p_{\text{rep}} = \mathcal{Y}_{\text{rep}}\mathcal{O}^m\,,
\ee
where $\mathcal{Y}_{\text{rep}}$ is the idempotent element of $S_n$'s irreducible representation rep, which are symbolised by Young diagrams.

Let us illustrate the procedure above by the type ${Q_L}{Q^\dagger_R}{B_R}{\phi^2}D^2$ at the NNLO. There are 5 fields in this type, and their helicities are $\{-1/2,-1/2,1,0,0\}$, and the last two NGBs, marked by indices 4 and 5, are repeated fields. There are 2 derivatives, thus $n_D = 2$. According to Tab~\ref{tab:buildingblocks}, we obtain that $h=0, r=4$, and the numbers of $\epsilon,\tilde{\epsilon}$ are $n=\tilde{n}=2$. Thus the primary Young diagram takes the form that
\be
    \yng(4,4,2)\,.
\ee
To fill this diagram and obtain SSYTs, the numbers of all indices are needed, which can be obtaind by Eq.~\ref{eq:numi}
\be
\#1 = 3,\quad \#2 = 3,\quad \#3=0,\quad \#4=2,\quad \#5=2\,,
\ee
thus there are only 4 SSYTs,
\be
    \young(1114,2225,45)\,,\quad \young(1112,2255,44)\,,\quad \young(1112,2244,55)\,,\quad \young(1112,2245,45)\,,
\ee
and they correspond to the Lorentz y-basis
\bea
\mathcal{B}^y_1 &=& -\lra{12}\lra{45}\lrs{34}\lrs{35}\,, \notag \\
\mathcal{B}^y_2 &=& \lra{15}\lra{25}\lrs{35}^2\,, \notag \\
\mathcal{B}^y_3 &=& \lra{14}\lra{24}\lrs{34}^2\,, \notag \\
\mathcal{B}^y_4 &=& -\lra{14}\lra{25}\lrs{34}\lrs{3,5}\,.
\eea
The Adler zero condition constrains that the amplitudes should be zero whenever the particle 4 or 5 becomes soft. If the particle 4 becomes soft, all the four basis becomes zero except for the second one $\mathcal{B}^y_2$, thus the matrix $\mathcal{K}$ in Eq.~\ref{eq:kil} takes the form that
\be
\mathcal{K} = \left(
\begin{array}{cccc}
    0 & 0 & 0 & 0 \\
    0 & 1 & 0 & 0 \\
    0 & 0 & 0 & 0 \\
    0 & 0 & 0 & 0 
\end{array}
\right)\,,
\ee
thus system of the linear equations in Eq.~\ref{eq:sysoflin} gives the solution that $c_2=0$. Similarly, if the particle 5 becomes soft, only the third basis $\mathcal{B}^y_3$ is not zero, thus there is the solution $c_3=0$. Thus only the first and the last bases satisfy the Adler zero condition, thus the actual Lorentz y-basis is 
\be
\mathcal{B}^{ay}_1 = \mathcal{B}^y_1,\quad \mathcal{B}^{ay}_2 = \mathcal{B}^y_4\,,
\ee
with the projection matrix that
\be
\mathcal{K}^{ay}_y = \left(\begin{array}{cccc} 1&0&0&0\\0&0&0&1\end{array}\right)\,.
\ee
According to Tab~\ref{tab:buildingblocks}, we translate these amplitudes to the form of operators
\bea
\mathcal{B}^{ay}_1 &=& {\psi_1}^\alpha{\psi_2}_\alpha {F_{R3}}_{\dot{\beta}\dot{\gamma}}(D^{\delta\dot{\beta}}\phi_4)(D_\delta^{\dot{\gamma}}\phi_5) \notag \\
&=& -4(\psi_1\psi_2){F_{R3}}^{\mu\nu}(D_\mu\phi_4)(D_\nu\phi_5)\,, \\
\mathcal{B}^{ay}_2 &=& {\psi_1}^\alpha{\psi_2}^\beta{R_{R3}}_{\dot{\gamma}\dot{\delta}}(D_\alpha^{\dot{\gamma}}\phi_4)(D_\beta^{\dot{\delta}}\phi_5) \notag \\
&=& -2(\psi_1\psi_2){F_{R3}}^{\mu\nu}(D_\mu\phi_4)(D_\nu\phi_5) +2i(\psi_1\sigma_{\mu\nu}\psi_2){F_{R3}}^{\nu\lambda}(D_\lambda\phi_4)(D^\mu\phi_5)\,,
\eea
from which we choose the monomials
\bea
\mathcal{B}^m_1 &=&(\psi_1\psi_2){F_{R3}}^{\mu\nu}(D_\mu\phi_4)(D_\nu\phi_5)\,, \\ 
\mathcal{B}^m_2 &=& (\psi_1\sigma_{\mu\nu}\psi_2){F_{R3}}^{\nu\lambda}(D_\lambda\phi_4)(D^\mu\phi_5) \,,
\eea
as the Lorentz m-basis, with the transformation matrix 
\be
\mathcal{K}^m_{ay} = \left(\begin{array}{cc}-\frac{1}{4}&0\\\frac{i}{4}&-\frac{i}{2}\end{array}\right)\,.
\ee

As for the gauge y-basis, the $SU(3)_C$ structure is simple, there is only one independent tensor $\delta_a^b$, and the $SU(2)_L$ tensors can be obtained by similar SSYT technics. There are 2 gauge SSYTs for this type
\be
\young(\ione\itwo\ifour,\jfour\ifive\jfive),\quad \young(\ione\ifour\jfour,\itwo\ifive\jfive)\,.
\ee
They correspond to the independent tensors 
\bea
\mathcal{T}^{(y)}_{SU(2)_L,1} &=& \epsilon^{i_1j_4}\epsilon^{i_2i_5}\epsilon^{i_4j_5}(\tau^{I_4})_{i_4i_4}(\tau^{I_5})_{i_5j_5}\epsilon_{i_2j_2}\,, \\
\mathcal{T}^{(y)}_{SU(2)_L,2} &=& \epsilon^{i_1i_2}\epsilon^{i_4i_5}\epsilon^{j_4j_5}(\tau^{I_4})_{i_4j_4}(\tau^{I_5})_{i_5j_5}\epsilon_{i_2j_2}\,, 
\eea
which can be further simplified to 
\bea
\mathcal{T}^{(y)}_{SU(2)_L,1} &=& \delta^{I_4I_5}\delta^{i_1}_{j_2} - i\epsilon^{I_4I_5J}(\tau^J)^{i_1}_{j_2}\,, \notag \\
\mathcal{T}^{(y)}_{SU(2)_L,2} &=& 2\delta^{I_4I_5}\delta^{i_1}_{j_2}\,.
\eea
Similarly, we combine them to obtain the gauge m-basis 
\be
\mathcal{T}^{(m)}_{SU(2)_L,1} = \epsilon^{I_4I_5J}(\tau^J)^{i_1}_{j_2},\quad \mathcal{T}^{(m)}_{SU(2)_L,2} = \delta^{I_4I_5}\delta^{i_1}_{j_2}\,,
\ee
with the transformation matrix 
\be
\mathcal{K}^{my}_{SU(2)_L} = \left(\begin{array}{cc}i&-\frac{i}{2}\\0&\frac{1}{2}\end{array}\right)\,.
\ee
The tensor product of the gauge m-basis and the Lorentz m-basis gives $1\times 2\times 2=4$ operators, but they are not all physical when considering the repeated fields. 

The NGB $\phi$ is the repeated field in this type, which carry no flavor number, thus only the operators symmetric under the permutations of them are physical. In the Lorentz space, the generator of $S_2$ in the y-basis~\footnote{The 2 generators of $S_2$ are identical, thus we treat them as one here.} $\{\mathcal{B}^y_i,i=1,2,3,4\}$ takes the form
\be
\mathcal{D}_\mathcal{B}^{(y)} = \left(\begin{array}{cccc}-1&0&0&0\\0&0&1&0\\0&1&0&0\\0&0&0&1\end{array}\right)\,,
\ee
which is obtained directly from manipulations of the amplitudes in the Lorentz y-basis. When focusing on the subspace spanned $\{\mathcal{B}^{m}_i,i=1,2\}$, the generator can be obtained by the linear transformation 
\be
\mathcal{D}_\mathcal{B}^{(m)} = \mathcal{K}^m_{ay}\mathcal{K}^{ay}_y \mathcal{D}_\mathcal{B}^{(y)} {\mathcal{K}^{ay}_y}^{-1} {\mathcal{K}^m_{ay}}^{-1} = \left(\begin{array}{cc}-1&0\\i&-2i\end{array}\right)\,,
\ee
where the matrix $\mathcal{K}^{ay}_{y}$ in the expression should be understood as
\be
\mathcal{K}^{ay}_y \rightarrow \left(\begin{array}{cccc}
    1 & 0 & 0 & 0 \\
    0 & 0 & 0 & 0 \\
    0 & 0 & 0 & 0 \\
    0 & 0 & 0 & 1
\end{array}\right)\,,
\ee
and the $2\times 2$ generators obtained above is the $4\times 4$ one with removal of all the null rows and columns.
In the case of the $SU(2)_L$ gauge basis, it is straightforward to obtain the generator of $S_2$ in the m-basis is 
\be
\mathcal{D}^{(m)}_{SU(2)_L} = \left(\begin{array}{cc}-1&0\\0&1\end{array}\right)\,,
\ee
and $\mathcal{D}^{(m)}_{SU(3)_C} = 1$.
Thus the generator of the overal operator space is the tensor product of them,
\be
\mathcal{D}^{(m)} = \mathcal{D}^{(m)}_{SU(3)_C}\otimes\mathcal{D}^{(m)}_{SU(2)_L}  \otimes \mathcal{D}_\mathcal{B}^{(m)} = \left(\begin{array}{cccc}-1&0&0&0\\2i&0&0&0\\0&0&-1&0\\0&0&2i&0\end{array}\right)\,.
\ee
The idempotent element of subspace is symmetric under the permutations of the 2 $\phi$'s, which is symbolised by the young diagram {\tiny\yng(2)}
\be
\caly[{\tiny\young(45)}] = I_{4\times 4}+\mathcal{D}^{(m)} = \left(\begin{array}{cccc}0&0&0&0\\2i&1&0&0\\0&0&0&0\\0&0&2i&1\end{array}\right)\,,
\ee
whose rank is 2, thus there are only 2 independent operators in this type, and we write them as following
\bea
\mathcal{O}_1 &=& \caly[{\tiny\young(45)}]\epsilon^{IJK}{\tau^K}^i_j({Q_L}_{ai}\sigma_{\mu\nu}{Q^\dagger_R}^{aj}){B_R}^{\nu\lambda}(D_\lambda\phi^I)(D^\mu\phi^J)\,,\notag \\
\mathcal{O}_2 &=& \caly[{\tiny\young(45)}]({Q_L}_{ai}\sigma_{\mu\nu}{Q^\dagger_R}^{ai}){B_R}^{\nu\lambda}(D_\lambda\phi^I)(D^\mu\phi^I)\,.
\eea
In the rest of this work, we will always write operators with the idempotent elements to indicate their permutation symmetry, but for the operators with no repeated fields and/or the repeated fields carrying flavor number 1, the idempotent elements will be omitted.

Another non-trivial example is the type ${B_L\phi^3hD^4}$ in the bosonic sector, with the helicity structure $\{-1,0,0,0,0\}$, where the particles $2\sim 4$ are the NGBs, the repeated fields, and the last one is the physical Higgs $h$. The $SU(3)_C$ gauge structure is trivial, and the $SU(2)_L$ gauge space is of dimension 1, $\epsilon^{I_2I_3I_4}$. The Lorentz basis of this type is complicated, since there are 4 derivatives. The filling of the SSYTs gives that the Lorentz space is of dimension 14, but the Adler zero condition constrains this space to the one of dimension 8, which has been discussed previously and the 8 basis are presented in Eq.~\ref{eq:examp2}. Combining the Lorentz and gauge structures together, we can obtain the f-basis is of dimension 1, and thus there is only 1 operator in this type
\be
\epsilon^{IJK}{B_L}^{\lambda\nu}(D^\mu h)(D_\mu\phi^I)(D_\nu\phi^J)(D_\lambda\phi^K)\,.
\ee
To be consistent with the building blocks of the HEFT, we replace the fields such as $(D_\mu\phi^I)$ in the operators obtained by the Young tensor method by the $\bfv^I_\mu$. Thus the only operator in this type is written as
\be
\epsilon^{IJK}{B_L}^{\lambda\nu}(D^\mu h)(\bfv_\mu^I)(\bfv_\nu^J)(\bfv_\lambda^K)\,.
\ee
This convention is respected in the section~\ref{sec:next}.

Next, let us provide some examples involving the spurions. Let us consider the type $W_L{L_L}{Q_L}{L_L^\dagger}{Q_L^\dagger}\bft^2$ from the class $\psi^4XUh$. There are 2 spurions in this type, because they do not participate in the Lorentz structures, only the particles $\{W_L,L_L,Q_L,L_L^\dagger,Q_L^\dagger\}$, corresponding to helicities $\{-1,-\frac{1}{2},-\frac{1}{2},\frac{1}{2},\frac{1}{2}\}$ are considered in the Lorentz basis. The Lorentz basis of this type is simple, and there is only one y-basis
\be
\mathcal{B}^{(y)}_1=\lra{12}\lra{13}\lrs{45}
\ee
corresponding to the m-basis
\be
{W_L}_{\mu\nu}({L_L}\sigma^\nu{L_L^\dagger})({Q_L}\sigma^\mu{Q_L^\dagger})
\ee
with the trivial transformation matrix.

For the gauge basis, the $SU(3)_C$ gauge basis is just $\delta^a_b$, contracting with the quarks $Q_L\,,Q_L^\dagger$, but the $SU(2)_L$ structures are complicated. There are 13 independent tensors in the $SU(2)_L$ gauge m-basis, but most of them are redundant. To eliminate those, we consider the spurions and the other fields separately. The 2 spurions takes the representation that
\begin{equation}
    \young(\isix\jsix\iseven\jseven) \label{eq:gauge_y1}
\end{equation}
according to the discussion around Eq.~\ref{eq:spinj}, where all the indices are symmetric. Its compensation takes the form that
\begin{equation}
\label{eq:shape}
\yng(5,1)\,,
\end{equation}
thus we need to construct all the Young diagrams of such shape by the outer product from the dynamical fields except the spurions, whose representations are related to the $SU(2)$ irreps
\begin{align}
    {W^{I_1}_L}{\tau^{{I_1}k}}_{i_1}\epsilon_{kj_1} & \sim \young(\ione\jone)\notag \\
    {L_L}_{i_2} &\sim \young(\itwo)\notag \\
    {Q_L}_{i_3} &\sim \young(\ithree)\notag \\
    {L^\dagger_L}^l\epsilon_{l i_4} &\sim \young(\ifour) \notag \\
    {Q^\dagger_L}^m\epsilon_{mi_5} &\sim \young(\ifive)\,,
\end{align}
and the outer product of them can be determined by the LR rule, 
\begin{align}
&\young(\ione\jone)\otimes\young(\itwo)\otimes\young(\ithree)\otimes\young(\ifour)\otimes\young(\ifive) \notag \\
    &=\left(\young(\ione\jone\itwo)\oplus\young(\ione\jone,\jtwo)\right)\otimes\young(\ithree)\otimes\young(\ifour)\otimes\young(\ifive) \notag \\
    &= \left(\young(\ione\jone\itwo\ithree)\oplus\young(\ione\jone\itwo,\ithree)\oplus\young(\ione\jone\ithree,\jtwo)\right)\otimes\young(\ifour)\otimes\young(\ifive) \notag \\
    &= \left(\young(\ione\jone\itwo\ithree\ifour)\oplus\young(\ione\jone\itwo\ithree,\ifour)\oplus\young(\ione\jone\itwo\ifour,\ithree)\oplus\young(\ione\jone\ithree\ifour,\jtwo)\right)\otimes\young(\ifive) \notag \\
    &= \young(\ione\jone\itwo\ithree\ifour,\ifive)\oplus\young(\ione\jone\itwo\ithree\ifive,\ifour)\oplus\young(\ione\jone\itwo\ifour\ifive,\ithree)\oplus\young(\ione\jone\ithree\ifour\ifive,\jtwo)\,, \label{eq:gauge_y2}
\end{align}
where only the Young diagrams as the same as the one in Eq.~\ref{eq:shape} is reserved.
Combining the diagrams \ref{eq:gauge_y1} and \ref{eq:gauge_y2} together, we obtain the gauge singlet representations with all the redundancies of the spurions eliminated:
\begin{align}
\mathcal{B}^{(y)}_{SU(2)_L,1} &= \young(\ione\jone\itwo\ithree\ifour,\ifive\isix\jsix\iseven\jseven) = {\tau^{I_1i_6}}_{i_1}\epsilon^{i_1i_5}(\bft^{I_6}{\tau^{I_6i_2}}_{i_6})(\bft^{I_7}{\tau^{I_7i_4}}_{i_7}\epsilon^{i_3i_7})\,,\notag \\
\mathcal{B}^{(y)}_{SU(2)_L,2} &= \young(\ione\jone\itwo\ithree\ifive,\ifour\isix\jsix\iseven\jseven) = {\tau^{I_1i_6}}_{i_1}\epsilon^{i_1i_4}(\bft^{I_6}{\tau^{I_6i_2}}_{i_6})(\bft^{I_7}{\tau^{I_7i_5}}_{i_7}\epsilon^{i_3i_7})\,,\notag \\
\mathcal{B}^{(y)}_{SU(2)_L,3} &= \young(\ione\jone\itwo\ifour\ifive,\ithree\isix\jsix\iseven\jseven) = {\tau^{I_1i_6}}_{i_1}\epsilon^{i_1i_3}(\bft^{I_6}{\tau^{I_6i_2}}_{i_6})(\bft^{I_7}{\tau^{I_7i_5}}_{i_7}\epsilon^{i_4i_7})\,,\notag \\
\mathcal{B}^{(y)}_{SU(2)_L,4} &= \young(\ione\jone\ithree\ifour\ifive,\itwo\isix\jsix\iseven\jseven) = {\tau^{I_1i_6}}_{i_1}\epsilon^{i_1i_2}(\bft^{I_6}{\tau^{I_6i_3}}_{i_6})(\bft^{I_7}{\tau^{I_7i_5}}_{i_7}\epsilon^{i_4i_7})\,.
\end{align}
It should be emphasized that each tensor above can be further simplified, for example, the first tensor takes the form
\begin{align}
    \mathcal{B}^{(y)}_{SU(2)_L,1} &= \bft^{I_6}\bft^{I_7}{\tau^{I_1i_6}}_{i_1}{\tau^{I_6i_2}}_{i_6}{\tau^{I_7i_4}}_{i_7}\epsilon^{i_1i_5}\epsilon^{i_3i_7} \notag \\
    &= \bft^{I_6}\bft^{I_7}{\tau^{I_1i_6}}_{i_1}{\tau^{I_6i_2}}_{i_6}{\tau^{I_7i_4}}_{i_7}(\delta^{i_1i_3}\delta^{i_5i_7}-\delta^{i_1i_7}\delta^{i_5i_3}) \notag \\
    &= \bft^{I_6}\bft^{I_7}{\tau^{I_1i_6}}_{i_3}{\tau^{I_6i_2}}_{i_6}{\tau^{I_7i_4}}_{i_5} - \bft^{I_6}\bft^{I_7}{\tau^{I_1i_6}}_{i_7}{\tau^{I_6i_2}}_{i_6}{\tau^{I_7i_4}}_{i_7}\delta^{i_3}_{i_5} \notag \\
    &= \bft^{I_6}\bft^{I_7}\epsilon^{I_6I_1K}{\tau^{Ki_2}}_{i_3}{\tau^{I_7i_4}}_{i_5} - \frac{1}{2}\bft^{I_6}\bft^{I_6}{\tau^{I_1i_2}}_{i_4}\delta_{i_5}^{i_3} + \frac{1}{2}\bft^{I_1}\bft^{{I_6}}{\tau^{I_6i_2}}_{i_4}\delta_{i_5}^{i_3} \,,
\end{align}
thus all the $SU(2)_L$ structures are usually polynomials, consistent with the method in the Ref~\cite{Sun:2022ssa}, which use the gauge j-basis techniques instead.

Thus there are 4 operators in this type, and we choose them as
\begin{align}
& i{{\tau^{K}}_{l}^{i}}{{\tau^{M}}_{k}^{j}}{\epsilon^{IJM}}{{\mathbf{T}}^{J}}{{\mathbf{T}}^{K}}{{{W_L}^{I}}_{\mu\nu}}({{{L_L}_{p}}_{i}}{\sigma^{\nu}}{{{L^\dagger_L}_{s}}^{k}})({{{Q_L}_{r}}_{aj}}\sigma^\mu{{{Q^\dagger_L}_{t}}^{al}})\,, \notag \\
& i{{\tau^{J}}_{l}^{j}}{{\mathbf{T}}^{I}}{{\mathbf{T}}^{J}}{{{W_L}^{I}}_{\mu\nu}}({{{L_L}_{p}}_{i}}{\sigma^{\nu}}{{{L^\dagger_L}_{s}}^{i}})({{{Q_L}_{r}}_{aj}}\sigma^\mu{{{Q^\dagger_L}_{t}}^{al}})\,, \notag \\
& i{{\tau^{J}}_{k}^{i}}{{\mathbf{T}}^{I}}{{\mathbf{T}}^{J}}{{{W_L}^{I}}_{\mu\nu}}({{{L_L}_{p}}_{i}}{\sigma^{\nu}}{{{L^\dagger_L}_{s}}^{k}})({{{Q_L}_{r}}_{aj}}\sigma^\mu{{{Q^\dagger_L}_{t}}^{aj}})\,, \notag \\
& i{{\tau^{J}}_{l}^{i}}{{\mathbf{T}}^{I}}{{\mathbf{T}}^{J}}{{{W_L}^{I}}_{\mu\nu}}({{{L_L}_{p}}_{i}}{\sigma^{\nu}}{{{L^\dagger_L}_{s}}^{j}})({{{Q_L}_{r}}_{aj}}\sigma^\mu{{{Q^\dagger_L}_{t}}^{al}})\,,
\end{align}
where the idempotent elements are omitted since the repeated fields $\bft$ have the flavor number 1 and are symmetric under permutations. 

Furthermore, if we consider the type with one more spurion, $W_L{L_L}{Q_L}{L_L^\dagger}{Q_L^\dagger}\bft^3$, the Lorentz basis is unchanged while the $SU(2)_L$ gauge space become dimension-30. The 3 spurions form the total symmetric representation
\begin{equation}
    \yng(6)\,,
\end{equation}
and the compensation takes the same shape
\begin{equation}
    \yng(6)\,.
\end{equation}
According to the LR rule, there is only one way to form such diagram from the outer product of the other fields, thus there is only one operator, though there are many spirions in this type, 
\be
{{\tau^{J}}_{k}^{i}}{{\tau^{K}}_{l}^{j}}{{\mathbf{T}}^{I}}{{\mathbf{T}}^{J}}{{\mathbf{T}}^{K}}{{{W_L}^{I}}_{\mu\nu}}({{{L_L}_{p}}_{i}}{\sigma^{\nu}}{{{L^\dagger_L}_{s}}^{k}})({{{Q_L}_{r}}_{aj}}\sigma^\mu{{{Q^\dagger_L}_{t}}^{al}})\,.
\ee
If the number of spurions in this type is more than 3, it is impossible to combine the totally symmetric combinations of spurions and other building blocks to form the gauge singlet, and thus there are no independent operator for this type.


Still there are cases that we have to consider many spurions in some types such as $\phi^6D^6\bft^6$. Though the $SU(2)_L$ gauge basis of this type is of high-dimension, there is only one independent operator according to the similar argument with the previous example, where the Young diagrams of the spurions is of the same shape with its compensation.

Finally, it should be emphasised again that, the $SU(2)_L$ is special and simple. If the spurions belong to the adjoint representation under the $SU(N)\,,N>2$ group, it is more difficult to deal with, since the corresponding Young diagrams are complicated.

\section{Next-to-next-to-leading-order Lagrangian}
\label{sec:next}

\subsection{Numbers of Operators at NNLO}
There are 12 classes of operators in the NNLO Lagrangian, ranging from chiral dimension ${\mathcal O}(p^5)$ and ${\mathcal O}(p^6)$, and numbers of the NNLO operators are listed in the Tab~\ref{tab:nnlonumb}. The explicit operators are presented as follows, with several comments in order.

\begin{table}[]
    \centering
    \begin{tabular}{c|c|c}
    \hline
 Class & $\mathcal{N}_{\text{term}}$ & $\mathcal{N}_{\text{operator}}$  \\
 \hline
$UhD^6$  &  114 & 114 \\
\hline
$X^2UhD^2$ & 130 & 130 \\
\hline
$XUhD^4$ & 164 & 164 \\
\hline
$\psi^2XUhD$ & 184 & $184{n_f}^2$ \\
\hline
$\psi^2UhX^2$ & 192 & $192{n_f}^2$ \\
\hline
$\psi^4UhD$ & 1224 & $\frac{4}{3}{n_f}^2(-2+3{n_f}+575{n_f}^2)$ \\
\hline
$\psi^4XUh$ & 1988 & $2{n_f}^2(-9+4n_f+519{n_f}^2)$ \\
\hline
$\psi^2UhD^3$ & 272 & $272{n_f}^2$ \\
\hline
$\psi^2XUhD^2$ & 1044 & $1044{n_f}^2$ \\
\hline
$\psi^4UhD^2$ & 8260 & $\frac{1}{3}{n_f}^2(155+78{n_f}+13525{n_f}^2)$ \\
\hline
$\psi^6Uh$ & 7112 & $\frac{1}{9}{n_f}^2(32+78{n_f}-133{n_f}^2+102{nf}^3+17129{n_f}^4)$ \\
\hline
$\psi^2UhD^4$ & 1112 & $1112{n_f}^2$ \\
\hline
\multirow{2}{*}{Total} & \multirow{2}{*}{21796} & $\frac{1}{9}(3672+25547{n_f}^2+420{n_f}^3+56684{n_f}^4+102{n_f}^5+17129{n_f}^6)$ \\
& &  $n_f=1$: 11506, $n_f=3$: 1927574 \\
\hline
    \end{tabular}
    \caption{The numbers of the NNLO operators in each class.}
    \label{tab:nnlonumb}
\end{table}

\bit

\item We adopt the weyl spionrs $\psi_{L/R}$ and $\psi^\dagger_{L/R}$ representing the SM fermions. The relations between the Weyl and the Dirac spinors $\Psi_{L/R}$ are
\be
\Psi_L = \left(\begin{array}{c}\psi_L\\0\end{array}\right),\quad \Psi_R = \left(\begin{array}{c}0\\\psi_R\end{array}\right),\quad \bar{\Psi}_L = \left(\begin{array}{c}0\\\psi^\dagger_L\end{array}\right),\quad \bar{\Psi}_R = \left(\begin{array}{c}\psi^\dagger_R\\0\end{array}\right)\,.
\ee
The detailed transformation between these two different notations can be found in Ref~\cite{Li:2020xlh}. Besides, the fermions contain the following $SU(3),SU(2)$ and flavor indices. We adopt the indices set of them as
\begin{align}
    SU(3)_{\rm color} &\sim \{a,b,c,d,e,f\} \notag \\
    SU(2)_{\rm L} &\sim \{i,j,k,l,m,n\} \notag \\
    SU(3)_{\rm flavor} &\sim \{p,r,s,t,u,v\}\,,
\end{align}
Under this convention, the quark doublet field is denoted as 
\begin{equation}
    {{Q_L}_p}_{ai}\,,\quad {{Q_L^\dagger}_r}^{ai}\,,\quad {{Q_R}_p}_{ai}\,,\quad {{Q^\dagger_R}_r}^{ai}\,,
\end{equation}
since the left- and right-handed spinors belong to fundamental representations while their hermitian conjugates belong to anti-fundamental representations, as shown in Tab~\ref{tab:buildingblocks}. In particular, the spinor indices of the spinor fields are not written explicitly, and are regarded as contracted with the neighbor spinor fields. For example, the bilinears should be understood as
\begin{align}
    (Q_L Q^\dagger_R) &= ({Q_L}^\alpha {Q^\dagger_R}_\alpha) \notag \\
    (Q_L \sigma^\mu Q_L^\dagger) &= ({Q_L}^\alpha \sigma^\mu_{\alpha\dot{\beta}}{Q_L^\dagger}^{\dot{\beta}}) \,.
\end{align}
\item The operators listed below assumes there are the right-handed neutrinos in the right-handed fermion doublet. If one removes the right-handed neutrinos in the fermion doublet, some operators involving in the spurion fields would disappear, because combination of the the spurion $\bft$ and the identity matrix would become the projector that picks up the right-handed neutrinos and the right-handed electrons. In this case, let us redefine the building blocks of the HEFT as the combination of the spurion $\bft$ and the identity matrix
\begin{align}
    \bft_+ &= \frac{I}{2} + \bft = \bfu(\frac{I}{2}+\mathcal{T}_R)\bfu^\dagger = \bfu\left(\begin{array}{cc}
       1  &  0\\
       0  &  0
    \end{array}\right)\bfu^\dagger \,,\notag \\
    \bft_- &= \frac{I}{2} - \bft = \bfu(\frac{I}{2}-\mathcal{T}_R)\bfu^\dagger = \bfu\left(\begin{array}{cc}
       0  &  0\\
       0  &  1
    \end{array}\right)\bfu^\dagger \,,
\end{align}
which can be regarded as another choice of the building blocks to characterise the custodial symmetry breaking.
The operators $\bft_+ $ and $\bft_- $ apply to the right-handed leptons would project out the right-handed neutrino $\nu_R$ and right-handed electron $e_R$
\begin{align}
    \bft_+L_R = \bfu\left(\begin{array}{cc}
       1  &  0\\
       0  &  0
    \end{array}\right)\left(\begin{array}{c}
    \nu_R  \\
    e_R 
    \end{array}\right) = \bfu\nu_R\,,\notag \\
    \bft_-L_R = \bfu\left(\begin{array}{cc}
       0  &  0\\
       0  &  1
    \end{array}\right)\left(\begin{array}{c}
    \nu_R  \\
    e_R 
    \end{array}\right) = \bfu e_R\,.\label{eq:tpm}
\end{align}
Thus in the case that the right-handed neutrinos are absent, the operators with $\bft_+$ acting on the right-handed lepton doublets should be eliminated. This becomes clear only when the operators presented below are re-expressed using the $\bft_+ $ and $\bft_- $, which are usually combinations of the original operators. For example, there are 3 operators (not all operators)
\begin{align}
    \calo_1 &= ({{{L_L}_{p}}_{i}}{{{L^\dagger_R}_{r}}^{i}}){{\bfv^I}_\mu}{{\bfv^I}^\mu}{{\bfv^K}_\nu}{{\bfv^K}^\nu}\,,\notag \\
    \calo_2 &= {{\tau^{O}}_{j}^{i}}{{\mathbf{T}}^{O}}({{{L_L}_{p}}_{i}}{{{L^\dagger_R}_{r}}^{j}}){{\bfv^I}_\mu}{{\bfv^I}^\mu}{{\bfv^J}_\nu}{{\bfv^J}^\nu} \,,\notag \\
    \calo_3 &= {\epsilon^{JKO}}{{\mathbf{T}}^{O}}({{{L_L}_{p}}_{i}}{\sigma^{\lambda\nu}}{{{L^\dagger_R}_{r}}^{i}}){{\bfv^I}_\lambda}{{\bfv^I}_\mu}{{\bfv^J}^\mu}{{\bfv^K}_\nu}
\end{align}
in the NNLO class $\psi^2UhD^4$. The combination of the first two operators $\calo_1/2+\calo_2$ involves the right-handed neutrinos $\nu_R$ according to Eq.~\ref{eq:tpm}, and should be eliminated if the $\nu_R$s are absent. While the third operator $\calo_3$ should be reserved, since the spurion $\bft$ the spurion field is contracted with the NGBs rather than the right-handed leptons. Thus in these 3 operators, there are only 2 of them are independent if the right-handed neutrinos are absent.

\item 
Every building block $\bfv_\mu$ in the class corresponds to pairing of the $\phi$ and $D$ in such type, while in the operators, the building block $\bfv_\mu$ instead of $D_\mu\phi$ is adopted.

\item The flavor structures of the operators are indicated by the idempotent elements before that, and the indices in the Young diagram are the flavor indices of the repeated fields, just like
\begin{equation}
    \mathcal{Y}[{\tiny\ytableaushort{p,r},\ytableaushort{st}}]{{\tau^{K}}_{l}^{i}}{\epsilon^{IJK}}({{{L_L}_{p}}_{i}}{{\sigma_\lambda}^{\mu}}{{{L_L}_{r}}_{j}})({{{L^\dagger_R}_{s}}^{j}}{\sigma^{\lambda\nu}}{{{L^\dagger_R}_{t}}^{l}}){{\mathbf{V}^I}_\mu}{{\mathbf{V}^J}_\nu}\,.
\end{equation}

\item The triple gauge bosons class $X^3Uh$, though carrying the chiral dimension 6, is attributed to the NLO operators, and is presented in the Ref.~\cite{Sun:2022ssa}, for the convenience of comparing with other literature.

\item The spurion is of the chiral dimension $d_\chi = 0$ to capture the possible custodial symmetry breaking effects at the LO Lagrangian, while in some literature~\cite{Pich:2016lew,Krause:2018cwe,Pich:2018ltt}, the spurion is taken to be of chiral dimension 1, then the NNLO Lagrangian would be the combination of the NLO operators listed in Ref.~\cite{Sun:2022ssa} and the NNLO operators listed below. 

\item In each class of operators, the physical Higgs singlet $h(x)$ could take arbitrary number in every operator, and thus we neglect the dimensionless Higgs function in each operator 
\bea
\mathcal{F} (h) = 1 + a \frac{h}{v} + b \frac{h^2}{v^2} + \cdots\,.
\eea

\item 
In some cases, certain powers of the Higgs field in the operator needs to be kept to be consistent with the derivatives on the Higgs field. In this case,  there is a minimum of the $h(x)$ fields for the independent operators. We utilize the convention that the number of Higgs $h$ in such type of operators keeps to be minimal to construct the complete and independent operators in this type.
 
\eit

%
%

Let us take the first type of operators $UhD^6$ as example to illustrate the above convention. In this type $UhD^6$, composed purely by Higgs $h$ and derivatives, the operators are listed as
\bea
h^2(D_\lambda D_\mu D_\nu h)(D^\lambda D^\mu D^\nu h), \quad 
h^3(D_\mu D_\nu h)(D_\lambda D^\mu h)(D^\lambda D^\nu h).
\eea
where the first operator can be constructed by 4 and 5 Higgses and 6 derivatives, but the second one can not be constructed until the number of Higgses is 6. 
This should be understood as the following. 
\bit
\item Suppose there are only 4 Higgses in the operator, we can only construct one single operator
\be
h^2(D_\lambda D_\mu D_\nu h)(D^\lambda D^\mu D^\nu h).
\ee
\item Suppose there are 5 Higgses in the operator, similarly we can construct one single operator
\be
h^3 (D_\lambda D_\mu D_\nu h)(D^\lambda D^\mu D^\nu h).
\ee
\item
Suppose the number of the Higgses are 6 and more, 
there are two kinds of operators with dimensionless Higgs function
\be
h^2(D_\lambda D_\mu D_\nu h)(D^\lambda D^\mu D^\nu h)\,\mathcal{F} (h) \,,\quad
h^3(D_\mu D_\nu h)(D_\lambda D^\mu h)(D^\lambda D^\nu h)\,\mathcal{F} (h) \,,
\ee

\eit
Since there are more than 6 Higgses, there is no more independent operator in this type, thus we write this type with the exact 6 Higgs and 6 derivatives, $h^6D^6$.


\subsection{Type: $UhD^6$}
\input{1_D6}
\subsection{Type: $X^2UhD^2$}
\input{2_X2D2}
\subsection{Type: $XUhD^4$}
\input{3_XD4}
\subsection{Type: $\psi^2XUhD$}
\input{4_psi2XD}
\subsection{Type: $\psi^2X^2Uh$}
\input{5_psi2X2}

\subsection{Type: $\psi^4UhD$}
\input{6_pis4D}

\subsection{Type: $\psi^4XUh$}
\input{7_psi4X}

\subsection{Type: $\psi^2UhD^3$}
\input{8_psi2D3}

\subsection{Type: $\psi^2XUhD^2$}
\input{9_psi2XD2}

\subsection{Type: $\psi^4UhD^2$}
\input{10_psi4D2_2}
\subsection{Type: $\psi^6Uh$}
\input{11_psi6_2}

\subsection{Type: $\psi^2UhD^4$}
\input{12_psi2D4}

\section{Conclusion}
\label{sec:conc}
In this work, we present the independent and complete NNLO operators of HEFT for the first time, by means of the Young tensor technique on the Lorentz, gauge and flavor structures, giving rise to the on-shell amplitude basis for each type of operators. For the operators involving in the Nambu-Goldstone bosons, the on-shell amplitude basis is further reduced to the subspace satisfying the Adler's zero condition in the soft momentum limit. 
The spurion field in the HEFT is carefully treated: they behave like ordinary building blocks with certain representation in the gauge sector, while we avoided the appearance of self-contractions among them. Thus the spurions can be implemented into the gauge structure by utilizing the Littlewood-Richardson rule on the symmetric Young diagram of the spurions and the ones of other dynamical fields. 
These new improvement on the NGBs and spurions in the Young tensor method are quite general and can thus be applied to other chiral effective field theories.

In the HEFT, we treat the power counting rules similar to the chiral Lagrangian, with the spurion field of no chiral dimension. We obtain that there are 11506 (1927574) NNLO operators, corresponding to the order ${\mathcal O}(p^5)$ and ${\mathcal O}(p^6)$, with one (three) generation fermion flavors. If the power counting rules treat the spurion field of chiral dimension one, then only a subset of the NNLO operators and partially the NLO operators with the spurion field consist of the complete and independent operators at the NNLO order.

Finally 
we expect the NNLO operators would be comparable with the one-loop corrections of the NLO operators in the HEFT. It also benefits the phenomenological studies of the dimension-8 standard model effective field theory in the broken phase. Especially when one performs the matching between the dimension-8 SMEFT and the NNLO HEFT operators, it is necessary to list the complete sets of the SMEFT and HEFT operators.


	
\section*{Acknowledgments} 
J.-H.Y. and H. S. are supported by the National Science Foundation of China under Grants No. 12022514, No. 11875003 and No. 12047503, and National Key Research and Development Program of China Grant No. 2020YFC2201501, No. 2021YFA0718304, and CAS Project for Young Scientists in Basic Research YSBR-006, the Key Research Program of the CAS Grant No. XDPB15. 
M.-L.X. is supported in part by the U.S. Department of Energy under contracts No. DE-AC02-06CH11357 at Argonne and No.DE-SC0010143 at Northwestern.

\bibliographystyle{JHEP}
\bibliography{ref}

\end{document}

%% file: 1_D6.tex
\begin{center}
\begin{longtable}{rl}
$ h^6D^6 $: & \\
\cline{1-1}
 & $h^2(D_\lambda D_\mu D_\nu h)(D^\lambda D^\mu D^\nu h)$ \\
& $h^3(D_\mu D_\nu h)(D_\lambda D^\mu h)(D^\lambda D^\nu h)$ \\
$\phi h^5\bft D^6$: & \\
\cline{1-1}
& $({D_\lambda}{D_\nu}{D_\mu}h)({D^\lambda}{D^\nu}h)h\mathbf{T}^I{{\bfv^I}^\mu}$ \\
& $({D_\lambda}{D_\nu}{D_\mu}h)({D^\nu}h)({D^\lambda}h)h\mathbf{T}^I{{\bfv^I}^\mu}$ \\
& $({D_\nu}{D_\mu}h)({D_\lambda}{D^\nu}h)({D^\lambda}h)h^2\mathbf{T}^I{{\bfv^I}^\mu}$ \\
$ h^4\phi^2D^6 $: & \\
\hline
\end{longtable}
\end{center}

%% file: 2_X2D2.tex
\begin{center}
\begin{longtable}{rl}
$ {B_L}^2h{\mathbf{T}}\phi $: \\
\cline{1-1}
& $ h{{\mathbf{T}}^{I}}{{B_L}^{\nu\lambda}}({D^{\mu}}{{B_L}_{\nu\lambda}})({D_{\mu}}{\phi^{I}}) $ \\
 
$ {{B_L}^2}{\phi^2} D^2$: \\
\cline{1-1}
& $ {{B_L}_{\nu\lambda}}{{B_L}^{\nu\lambda}}{{\mathbf{V}^I}_\mu}{{\mathbf{V}^I}^\mu} $ \\
 
$ {{B_L}^2}{\mathbf{T}}^2{\phi^2} D^2$: \\
\cline{1-1}
& $ {{\mathbf{T}}^{I}}{{\mathbf{T}}^{J}}{{B_L}_{\nu\lambda}}{{B_L}^{\nu\lambda}}{{\mathbf{V}^I}_\mu}{{\mathbf{V}^J}^\mu} $ \\
 
$ {B_L}{W_L}h\phi D^2$:\\
\cline{1-1}
& $ h{{{W_L}^{I}}^{\nu\lambda}}({D^{\mu}}{{B_L}_{\nu\lambda}}){{\mathbf{V}^I}_\mu} $ \\
\end{longtable}
\end{center}

%% file: 3_XD4.tex
\begin{center}
\begin{longtable}{rl}
 
$ {B_L}{\mathbf{T}}\phi^4 D^4$: \\
\cline{1-1}
& $  {\epsilon^{IKO}}{{\mathbf{T}}^{M}}{{B_L}^{\lambda\nu}}{{\bfv^I}_\mu}{{\bfv^K}_\nu}{{\bfv^M}_\lambda}{{\bfv^O}^\mu} $ \\
& $  {\epsilon^{JMN}}{{\mathbf{T}}^{N}}{{B_L}^{\lambda\nu}}{{\bfv^I}_\mu}{{\bfv^I}_\nu}{{\bfv^J}^\mu}{{\bfv^M}_\lambda} $ \\

$ {B_L}{\mathbf{T}}^2\phi^4 D^4$: \\
\cline{1-1}
& $  {{\mathbf{T}}^{J}}{{\mathbf{T}}^{K}}{{B_L}^{\lambda\nu}}{{\bfv^I}_\lambda}{{\bfv^I}_\mu}{{\bfv^J}_\nu}{{\bfv^K}^\mu} $ \\
 
$ {B_L}{\mathbf{T}}^3\phi^4 D^4 $: \\
\cline{1-1}
& $  {\epsilon^{KPQ}}{{\mathbf{T}}^{J}}{{\mathbf{T}}^{M}}{{\mathbf{T}}^{P}}{{B_L}^{\lambda\nu}}{{\bfv^J}^\mu}{{\bfv^K}_\nu}{{\bfv^M}_\lambda}{{\bfv^Q}_\mu} $ \\
 
$ {W_L}h\phi^3 D^4$: \\
\end{longtable}
\end{center}

%% file: 4_psi2XD.tex
\begin{center}
\begin{longtable}{rl}

${B_L}{Q_L}\phi{Q^\dagger_L} D$:\\
\cline{1-1}
& $\bfv^I_\mu({Q_L}_{pai}\sigma_\nu{\tau^I}^i_j{Q^\dagger_L}^{aj}_r) {B_L}^{\mu\nu}$ \\

${B_L}{Q_L}\phi{Q^\dagger_L}\bft D$: \\
\cline{1-1}
& $\bft^I\bfv^I_{\mu}({Q_L}_{pai}\sigma_\nu{Q^\dagger_L}^{ai}_r){B_L}^{\mu\nu}$\\
& $\epsilon^{IJK}\bft^J\bfv^I_\mu({Q_L}_{pai}\sigma_\nu{\tau^K}^i_j{Q^\dagger_L}^{aj}_r){B_L}^{\mu\nu}$\\

${B_L}{Q_L}\phi{Q^\dagger_L}\bft^2 D$: \\
\cline{1-1}
& $\bft^I\bft^J\bfv^I_\mu({Q_L}_{pai}\sigma_\nu{\tau^J}^i_j{Q^\dagger_L}^{aj}_r){B_L}^{\mu\nu}$\\

${G_L}{Q_L}\phi{Q^\dagger_L} D$: \\
\end{longtable}
\end{center}

%% file: 5_psi2X2.tex
\begin{center}
\begin{longtable}{rl}

$ {B_L}^2{Q^\dagger_L}{Q_R} $: \\ \cline{1-1} & $  {{B_L}_{\mu\nu}}{{B_L}^{\mu\nu}}({{{Q^\dagger_L}_{p}}^{ai}}{{{Q_R}_{r}}_{ai}}) $ \\

$ {B_L}^2{\mathbf{T}}{Q^\dagger_L}{Q_R} $: \\ \cline{1-1} & $  {{\tau^{I}}_{i}^{j}}{{\mathbf{T}}^{I}}{{B_L}_{\mu\nu}}{{B_L}^{\mu\nu}}({{{Q^\dagger_L}_{p}}^{ai}}{{{Q_R}_{r}}_{aj}}) $ \\

$ {B_L}{W_L}{Q^\dagger_L}{Q_R} $: \\ \cline{1-1} & $ {{\tau^{I}}_{i}^{j}}{{B_L}_{\mu\nu}}{{{W_L}^{I}}^{\mu\nu}}({{{Q^\dagger_L}_{p}}^{ai}}{{{Q_R}_{r}}_{aj}}) $ \\

$ {B_L}{W_L}{\mathbf{T}}{Q^\dagger_L}{Q_R} $: \\ \cline{1-1} & $ {{\tau^{K}}_{i}^{j}}{\epsilon^{IJK}}{{\mathbf{T}}^{J}}{{B_L}_{\mu\nu}}{{{W_L}^{I}}^{\mu\nu}}({{{Q^\dagger_L}_{p}}^{ai}}{{{Q_R}_{r}}_{aj}}) $ \\
\end{longtable}
\end{center}

%% file: 6_pis4D.tex
\begin{center}
\begin{longtable}{rl}

$ {L_L}{L^\dagger_R}^2\phi{L_R}D $: \\ \cline{1-1} & $ \mathcal{Y}[{\tiny\ytableaushort{rs}}]{{\tau^{I}}_{j}^{i}}({{{L_R}_{t}}_{k}}{\bar{\sigma}^{\mu}}{{{L_L}_{p}}_{i}})({{{L^\dagger_R}_{r}}^{j}}{{{L^\dagger_R}_{s}}^{k}}){{\mathbf{V}^I}_\mu} $ \\
 & $ \mathcal{Y}[{\tiny\ytableaushort{rs}}]{{\tau^{I}}_{j}^{i}}({{{L_R}_{t}}_{k}}{\bar{\sigma}_{\nu}}{{{L_L}_{p}}_{i}})({{{L^\dagger_R}_{r}}^{j}}{\sigma^{\nu\mu}}{{{L^\dagger_R}_{s}}^{k}}){{\mathbf{V}^I}_\mu} $ \\
 & $ \mathcal{Y}[{\tiny\ytableaushort{rs}}]{{\tau^{I}}_{j}^{l}}({{{L_R}_{t}}_{l}}{\bar{\sigma}^{\mu}}{{{L_L}_{p}}_{i}})({{{L^\dagger_R}_{r}}^{j}}{{{L^\dagger_R}_{s}}^{i}}){{\mathbf{V}^I}_\mu} $ \\
 & $ \mathcal{Y}[{\tiny\ytableaushort{r, s}}]{{\tau^{I}}_{j}^{i}}({{{L_R}_{t}}_{k}}{\bar{\sigma}^{\mu}}{{{L_L}_{p}}_{i}})({{{L^\dagger_R}_{r}}^{j}}{{{L^\dagger_R}_{s}}^{k}}){{\mathbf{V}^I}_\mu} $ \\
 & $ \mathcal{Y}[{\tiny\ytableaushort{r, s}}]{{\tau^{I}}_{j}^{i}}({{{L_R}_{t}}_{k}}{\bar{\sigma}_{\nu}}{{{L_L}_{p}}_{i}})({{{L^\dagger_R}_{r}}^{j}}{\sigma^{\nu\mu}}{{{L^\dagger_R}_{s}}^{k}}){{\mathbf{V}^I}_\mu} $ \\
 & $ \mathcal{Y}[{\tiny\ytableaushort{r, s}}]{{\tau^{I}}_{j}^{l}}({{{L_R}_{t}}_{l}}{\bar{\sigma}_{\nu}}{{{L_L}_{p}}_{i}})({{{L^\dagger_R}_{r}}^{j}}{\sigma^{\nu\mu}}{{{L^\dagger_R}_{s}}^{i}}){{\mathbf{V}^I}_\mu} $ \\

$ {L_L}{L^\dagger_R}^2{\mathbf{T}}\phi{L_R} D$: \\
\end{longtable}
\end{center}

%% file: 8_psi2D3.tex
\begin{center}
\begin{longtable}{rl}

$ {L^\dagger_R}{\phi}^3{L_R}D^3 $ : : \\ \cline{1-1} & $  {\epsilon^{IJK}}({{{L^\dagger_R}_{p}}^{i}}{\sigma_{\rho}}{{{L_R}_{r}}_{i}}){{\mathbf{V}^I}_\mu}{{\mathbf{V}^J}_\nu}{{\mathbf{V}^K}_\lambda}{\epsilon^{\lambda \mu \nu \rho}} $ \\
 & $  {{\tau^{J}}_{i}^{j}}({{{L^\dagger_R}_{p}}^{i}}{\sigma^{\nu}}{{{L_R}_{r}}_{j}}){{\mathbf{V}^I}_\nu}{{\mathbf{V}^I}^\mu}{{\mathbf{V}^J}_\mu} $ \\
 & $  {{\tau^{J}}_{i}^{j}}({{{L^\dagger_R}_{p}}^{i}}{\sigma^{\mu}}{{{L_R}_{r}}_{j}}){{\mathbf{V}^I}_\nu}{{\mathbf{V}^I}^\nu}{{\mathbf{V}^J}_\mu} $ \\

$ {L^\dagger_R}{\mathbf{T}}{\phi}^3{L_R}D^3 $ : : \\ \cline{1-1} & $  {{\tau^{K}}_{i}^{j}}{\epsilon^{IJM}}{{\mathbf{T}}^{K}}({{{L^\dagger_R}_{p}}^{i}}{\sigma_{\rho}}{{{L_R}_{r}}_{j}}){{\mathbf{V}^I}_\mu}{{\mathbf{V}^J}_\lambda}{{\mathbf{V}^M}_\nu}{\epsilon^{\lambda \mu \nu \rho}} $ \\
 & $  {{\tau^{J}}_{i}^{j}}{\epsilon^{IKM}}{{\mathbf{T}}^{M}}({{{L^\dagger_R}_{p}}^{i}}{\sigma^{\nu}}{{{L_R}_{r}}_{j}}){{\mathbf{V}^I}_\mu}{{\mathbf{V}^J}^\mu}{{\mathbf{V}^K}_\nu} $ \\
 & $  {{\tau^{M}}_{i}^{j}}{\epsilon^{JKM}}{{\mathbf{T}}^{K}}({{{L^\dagger_R}_{p}}^{i}}{\sigma^{\nu}}{{{L_R}_{r}}_{j}}){{\mathbf{V}^I}_\mu}{{\mathbf{V}^I}^\mu}{{\mathbf{V}^J}_\nu} $ \\
\end{longtable}
\end{center}

%% file: 9_psi2XD2.tex
\begin{center}
\begin{longtable}{rl}

$ {B_L}{L_L}{L^\dagger_R}{\phi}^2 D^2 $ : \\ \cline{1-1} & $  {{B_L}_{\nu\lambda}}({{{L_L}_{p}}_{i}}{\sigma^{\nu\lambda}}{{{L^\dagger_R}_{r}}^{i}}){{\mathbf{V}^I}_\mu}{{\mathbf{V}^I}^\mu} $ \\
 & $  {{\tau^{K}}_{j}^{i}}{\epsilon^{IJK}}{{B_L}^{\nu\mu}}({{{L_L}_{p}}_{i}}{{{L^\dagger_R}_{r}}^{j}}){{\mathbf{V}^I}_\mu}{{\mathbf{V}^J}_\nu} $ \\
 & $  {{\tau^{K}}_{j}^{i}}{\epsilon^{IJK}}{{{B_L}_{\lambda}}^{\mu}}({{{L_L}_{p}}_{i}}{\sigma^{\lambda\nu}}{{{L^\dagger_R}_{r}}^{j}}){{\mathbf{V}^I}_\mu}{{\mathbf{V}^J}_\nu} $ \\

$ {B_L}{L_L}{L^\dagger_R}{\mathbf{T}}{\phi}^2 D^2 $ : \\ \cline{1-1} & $  {\epsilon^{IJK}}{{\mathbf{T}}^{K}}{{B_L}^{\nu\mu}}({{{L_L}_{p}}_{i}}{{{L^\dagger_R}_{r}}^{i}}){{\mathbf{V}^I}_\mu}{{\mathbf{V}^J}_\nu} $ \\
 & $  {\epsilon^{IJK}}{{\mathbf{T}}^{K}}{{{B_L}_{\lambda}}^{\mu}}({{{L_L}_{p}}_{i}}{\sigma^{\lambda\nu}}{{{L^\dagger_R}_{r}}^{i}}){{\mathbf{V}^I}_\mu}{{\mathbf{V}^J}_\nu} $ \\
 & $  {{\tau^{J}}_{j}^{i}}{{\mathbf{T}}^{I}}{{B_L}^{\nu\mu}}({{{L_L}_{p}}_{i}}{{{L^\dagger_R}_{r}}^{j}}){{\mathbf{V}^I}_\nu}{{\mathbf{V}^J}_\mu} $ \\
\end{longtable}
\end{center}

%% file: 12_psi2D4.tex
\begin{center}
    \begin{longtable}{rl}

$ {L_L}{L^\dagger_R}{\phi^4} D^4 $ \\ \cline{1-1} & $ {{\tau^{K}}_{j}^{i}}{\epsilon^{IMN}}({{{L_L}_{p}}_{i}}{\sigma^{\lambda\nu}}{{{L^\dagger_R}_{r}}^{j}}){{\bfv^I}_\mu}{{\bfv^K}_\nu}{{\bfv^M}_\lambda}{{\bfv^N}^\mu} $ \\
 & $ {{\tau^{N}}_{j}^{i}}{\epsilon^{KMN}}({{{L_L}_{p}}_{i}}{\sigma^{\lambda\nu}}{{{L^\dagger_R}_{r}}^{j}}){{\bfv^I}_\mu}{{\bfv^I}_\nu}{{\bfv^K}^\mu}{{\bfv^M}_\lambda} $ \\
 & $ ({{{L_L}_{p}}_{i}}{{{L^\dagger_R}_{r}}^{i}}){{\bfv^I}_\mu}{{\bfv^I}^\mu}{{\bfv^K}_\nu}{{\bfv^K}^\nu} $ \\
 & $ ({{{L_L}_{p}}_{i}}{{{L^\dagger_R}_{r}}^{i}}){{\bfv^I}_\mu}{{\bfv^I}_\nu}{{\bfv^K}^\mu}{{\bfv^K}^\nu} $ \\

$ {L_L}{L^\dagger_R}{\mathbf{T}}{\phi^4} D^4 $ \\ \cline{1-1} & $ {\epsilon^{JKO}}{{\mathbf{T}}^{O}}({{{L_L}_{p}}_{i}}{\sigma^{\lambda\nu}}{{{L^\dagger_R}_{r}}^{i}}){{\bfv^I}_\lambda}{{\bfv^I}_\mu}{{\bfv^J}^\mu}{{\bfv^K}_\nu} $ \\
 & $ {\epsilon^{JKO}}{{\mathbf{T}}^{O}}({{{L_L}_{p}}_{i}}{\sigma^{\lambda\mu}}{{{L^\dagger_R}_{r}}^{i}}){{\bfv^I}_\lambda}{{\bfv^I}_\mu}{{\bfv^J}_\nu}{{\bfv^K}^\nu} $ \\
\end{longtable}
\end{center}

%% file: main.bbl
\providecommand{\href}[2]{#2}\begingroup\raggedright\begin{thebibliography}{10}

\bibitem{Weinberg:1978kz}
S.~Weinberg, \emph{{Phenomenological Lagrangians}},
  \href{http://dx.doi.org/10.1016/0378-4371(79)90223-1}{\emph{Physica A} {\bf
  96} (1979) 327--340}.

\bibitem{Weinberg:1979sa}
S.~Weinberg, \emph{{Baryon and Lepton Nonconserving Processes}},
  \href{http://dx.doi.org/10.1103/PhysRevLett.43.1566}{\emph{Phys. Rev. Lett.}
  {\bf 43} (1979) 1566--1570}.

\bibitem{Buchmuller:1985jz}
W.~Buchmuller and D.~Wyler, \emph{{Effective Lagrangian Analysis of New
  Interactions and Flavor Conservation}},
  \href{http://dx.doi.org/10.1016/0550-3213(86)90262-2}{\emph{Nucl. Phys. B}
  {\bf 268} (1986) 621--653}.

\bibitem{Grzadkowski:2010es}
B.~Grzadkowski, M.~Iskrzynski, M.~Misiak and J.~Rosiek, \emph{{Dimension-Six
  Terms in the Standard Model Lagrangian}},
  \href{http://dx.doi.org/10.1007/JHEP10(2010)085}{\emph{JHEP} {\bf 10} (2010)
  085}, [\href{https://arxiv.org/abs/1008.4884}{{\tt 1008.4884}}].

\bibitem{Lehman:2014jma}
L.~Lehman, \emph{{Extending the Standard Model Effective Field Theory with the
  Complete Set of Dimension-7 Operators}},
  \href{http://dx.doi.org/10.1103/PhysRevD.90.125023}{\emph{Phys. Rev. D} {\bf
  90} (2014) 125023}, [\href{https://arxiv.org/abs/1410.4193}{{\tt
  1410.4193}}].

\bibitem{Liao:2016hru}
Y.~Liao and X.-D. Ma, \emph{{Renormalization Group Evolution of Dimension-seven
  Baryon- and Lepton-number-violating Operators}},
  \href{http://dx.doi.org/10.1007/JHEP11(2016)043}{\emph{JHEP} {\bf 11} (2016)
  043}, [\href{https://arxiv.org/abs/1607.07309}{{\tt 1607.07309}}].

\bibitem{Li:2020gnx}
H.-L. Li, Z.~Ren, J.~Shu, M.-L. Xiao, J.-H. Yu and Y.-H. Zheng, \emph{{Complete
  set of dimension-eight operators in the standard model effective field
  theory}}, \href{http://dx.doi.org/10.1103/PhysRevD.104.015026}{\emph{Phys.
  Rev. D} {\bf 104} (2021) 015026},
  [\href{https://arxiv.org/abs/2005.00008}{{\tt 2005.00008}}].

\bibitem{Murphy:2020rsh}
C.~W. Murphy, \emph{{Dimension-8 operators in the Standard Model Eective Field
  Theory}}, \href{http://dx.doi.org/10.1007/JHEP10(2020)174}{\emph{JHEP} {\bf
  10} (2020) 174}, [\href{https://arxiv.org/abs/2005.00059}{{\tt 2005.00059}}].

\bibitem{Li:2020xlh}
H.-L. Li, Z.~Ren, M.-L. Xiao, J.-H. Yu and Y.-H. Zheng, \emph{{Complete set of
  dimension-nine operators in the standard model effective field theory}},
  \href{http://dx.doi.org/10.1103/PhysRevD.104.015025}{\emph{Phys. Rev. D} {\bf
  104} (2021) 015025}, [\href{https://arxiv.org/abs/2007.07899}{{\tt
  2007.07899}}].

\bibitem{Liao:2020jmn}
Y.~Liao and X.-D. Ma, \emph{{An explicit construction of the dimension-9
  operator basis in the standard model effective field theory}},
  \href{http://dx.doi.org/10.1007/JHEP11(2020)152}{\emph{JHEP} {\bf 11} (2020)
  152}, [\href{https://arxiv.org/abs/2007.08125}{{\tt 2007.08125}}].

\bibitem{Li:2020zfq}
H.-L. Li, J.~Shu, M.-L. Xiao and J.-H. Yu, \emph{{Depicting the Landscape of
  Generic Effective Field Theories}},
  \href{https://arxiv.org/abs/2012.11615}{{\tt 2012.11615}}.

\bibitem{Henning:2015alf}
B.~Henning, X.~Lu, T.~Melia and H.~Murayama, \emph{{2, 84, 30, 993, 560, 15456,
  11962, 261485, ...: Higher dimension operators in the SM EFT}},
  \href{http://dx.doi.org/10.1007/JHEP08(2017)016}{\emph{JHEP} {\bf 08} (2017)
  016}, [\href{https://arxiv.org/abs/1512.03433}{{\tt 1512.03433}}].

\bibitem{Li:2022tec}
H.-L. Li, Z.~Ren, M.-L. Xiao, J.-H. Yu and Y.-H. Zheng, \emph{{Operators for
  generic effective field theory at any dimension: on-shell amplitude basis
  construction}}, \href{http://dx.doi.org/10.1007/JHEP04(2022)140}{\emph{JHEP}
  {\bf 04} (2022) 140}, [\href{https://arxiv.org/abs/2201.04639}{{\tt
  2201.04639}}].

\bibitem{Coleman:1969sm}
S.~R. Coleman, J.~Wess and B.~Zumino, \emph{{Structure of phenomenological
  Lagrangians. 1.}},
  \href{http://dx.doi.org/10.1103/PhysRev.177.2239}{\emph{Phys. Rev.} {\bf 177}
  (1969) 2239--2247}.

\bibitem{Callan:1969sn}
C.~G. Callan, Jr., S.~R. Coleman, J.~Wess and B.~Zumino, \emph{{Structure of
  phenomenological Lagrangians. 2.}},
  \href{http://dx.doi.org/10.1103/PhysRev.177.2247}{\emph{Phys. Rev.} {\bf 177}
  (1969) 2247--2250}.

\bibitem{Appelquist:1980vg}
T.~Appelquist and C.~W. Bernard, \emph{{Strongly Interacting Higgs Bosons}},
  \href{http://dx.doi.org/10.1103/PhysRevD.22.200}{\emph{Phys. Rev. D} {\bf 22}
  (1980) 200}.

\bibitem{Longhitano:1980iz}
A.~C. Longhitano, \emph{{Heavy Higgs Bosons in the Weinberg-Salam Model}},
  \href{http://dx.doi.org/10.1103/PhysRevD.22.1166}{\emph{Phys. Rev. D} {\bf
  22} (1980) 1166}.

\bibitem{Longhitano:1980tm}
A.~C. Longhitano, \emph{{Low-Energy Impact of a Heavy Higgs Boson Sector}},
  \href{http://dx.doi.org/10.1016/0550-3213(81)90109-7}{\emph{Nucl. Phys. B}
  {\bf 188} (1981) 118--154}.

\bibitem{Feruglio:1992wf}
F.~Feruglio, \emph{{The Chiral approach to the electroweak interactions}},
  \href{http://dx.doi.org/10.1142/S0217751X93001946}{\emph{Int. J. Mod. Phys.
  A} {\bf 8} (1993) 4937--4972},
  [\href{https://arxiv.org/abs/hep-ph/9301281}{{\tt hep-ph/9301281}}].

\bibitem{Herrero:1993nc}
M.~J. Herrero and E.~Ruiz~Morales, \emph{{The Electroweak chiral Lagrangian for
  the Standard Model with a heavy Higgs}},
  \href{http://dx.doi.org/10.1016/0550-3213(94)90525-8}{\emph{Nucl. Phys. B}
  {\bf 418} (1994) 431--455}, [\href{https://arxiv.org/abs/hep-ph/9308276}{{\tt
  hep-ph/9308276}}].

\bibitem{Herrero:1994iu}
M.~J. Herrero and E.~Ruiz~Morales, \emph{{Nondecoupling effects of the SM higgs
  boson to one loop}},
  \href{http://dx.doi.org/10.1016/0550-3213(94)00589-7}{\emph{Nucl. Phys. B}
  {\bf 437} (1995) 319--355}, [\href{https://arxiv.org/abs/hep-ph/9411207}{{\tt
  hep-ph/9411207}}].

\bibitem{Buchalla:2012qq}
G.~Buchalla and O.~Cata, \emph{{Effective Theory of a Dynamically Broken
  Electroweak Standard Model at NLO}},
  \href{http://dx.doi.org/10.1007/JHEP07(2012)101}{\emph{JHEP} {\bf 07} (2012)
  101}, [\href{https://arxiv.org/abs/1203.6510}{{\tt 1203.6510}}].

\bibitem{Alonso:2012px}
R.~Alonso, M.~B. Gavela, L.~Merlo, S.~Rigolin and J.~Yepes, \emph{{The
  Effective Chiral Lagrangian for a Light Dynamical ''Higgs Particle''}},
  \href{http://dx.doi.org/10.1016/j.physletb.2013.04.037}{\emph{Phys. Lett. B}
  {\bf 722} (2013) 330--335}, [\href{https://arxiv.org/abs/1212.3305}{{\tt
  1212.3305}}].

\bibitem{Buchalla:2013rka}
G.~Buchalla, O.~Cat\`a and C.~Krause, \emph{{Complete Electroweak Chiral
  Lagrangian with a Light Higgs at NLO}},
  \href{http://dx.doi.org/10.1016/j.nuclphysb.2014.01.018}{\emph{Nucl. Phys. B}
  {\bf 880} (2014) 552--573}, [\href{https://arxiv.org/abs/1307.5017}{{\tt
  1307.5017}}].

\bibitem{Brivio:2013pma}
I.~Brivio, T.~Corbett, O.~J.~P. \'Eboli, M.~B. Gavela, J.~Gonzalez-Fraile,
  M.~C. Gonzalez-Garcia et~al., \emph{{Disentangling a dynamical Higgs}},
  \href{http://dx.doi.org/10.1007/JHEP03(2014)024}{\emph{JHEP} {\bf 03} (2014)
  024}, [\href{https://arxiv.org/abs/1311.1823}{{\tt 1311.1823}}].

\bibitem{Pich:2015kwa}
A.~Pich, I.~Rosell, J.~Santos and J.~J. Sanz-Cillero, \emph{{Low-energy signals
  of strongly-coupled electroweak symmetry-breaking scenarios}},
  \href{http://dx.doi.org/10.1103/PhysRevD.93.055041}{\emph{Phys. Rev. D} {\bf
  93} (2016) 055041}, [\href{https://arxiv.org/abs/1510.03114}{{\tt
  1510.03114}}].

\bibitem{Pich:2016lew}
A.~Pich, I.~Rosell, J.~Santos and J.~J. Sanz-Cillero, \emph{{Fingerprints of
  heavy scales in electroweak effective Lagrangians}},
  \href{http://dx.doi.org/10.1007/JHEP04(2017)012}{\emph{JHEP} {\bf 04} (2017)
  012}, [\href{https://arxiv.org/abs/1609.06659}{{\tt 1609.06659}}].

\bibitem{Brivio:2016fzo}
I.~Brivio, J.~Gonzalez-Fraile, M.~C. Gonzalez-Garcia and L.~Merlo, \emph{{The
  complete HEFT Lagrangian after the LHC Run I}},
  \href{http://dx.doi.org/10.1140/epjc/s10052-016-4211-9}{\emph{Eur. Phys. J.
  C} {\bf 76} (2016) 416}, [\href{https://arxiv.org/abs/1604.06801}{{\tt
  1604.06801}}].

\bibitem{Merlo:2016prs}
L.~Merlo, S.~Saa and M.~Sacrist\'an-Barbero, \emph{{Baryon Non-Invariant
  Couplings in Higgs Effective Field Theory}},
  \href{http://dx.doi.org/10.1140/epjc/s10052-017-4753-5}{\emph{Eur. Phys. J.
  C} {\bf 77} (2017) 185}, [\href{https://arxiv.org/abs/1612.04832}{{\tt
  1612.04832}}].

\bibitem{Krause:2018cwe}
C.~Krause, A.~Pich, I.~Rosell, J.~Santos and J.~J. Sanz-Cillero,
  \emph{{Colorful Imprints of Heavy States in the Electroweak Effective
  Theory}}, \href{http://dx.doi.org/10.1007/JHEP05(2019)092}{\emph{JHEP} {\bf
  05} (2019) 092}, [\href{https://arxiv.org/abs/1810.10544}{{\tt 1810.10544}}].

\bibitem{Falkowski:2019tft}
A.~Falkowski and R.~Rattazzi, \emph{{Which EFT}},
  \href{http://dx.doi.org/10.1007/JHEP10(2019)255}{\emph{JHEP} {\bf 10} (2019)
  255}, [\href{https://arxiv.org/abs/1902.05936}{{\tt 1902.05936}}].

\bibitem{Agrawal:2019bpm}
P.~Agrawal, D.~Saha, L.-X. Xu, J.-H. Yu and C.~P. Yuan, \emph{{Determining the
  shape of the Higgs potential at future colliders}},
  \href{http://dx.doi.org/10.1103/PhysRevD.101.075023}{\emph{Phys. Rev. D} {\bf
  101} (2020) 075023}, [\href{https://arxiv.org/abs/1907.02078}{{\tt
  1907.02078}}].

\bibitem{Cohen:2020xca}
T.~Cohen, N.~Craig, X.~Lu and D.~Sutherland, \emph{{Is SMEFT Enough?}},
  \href{http://dx.doi.org/10.1007/JHEP03(2021)237}{\emph{JHEP} {\bf 03} (2021)
  237}, [\href{https://arxiv.org/abs/2008.08597}{{\tt 2008.08597}}].

\bibitem{Sun:2022ssa}
H.~Sun, M.-L. Xiao and J.-H. Yu, \emph{{Complete NLO Operators in the Higgs
  Effective Field Theory}},  \href{https://arxiv.org/abs/2206.07722}{{\tt
  2206.07722}}.

\bibitem{Eboli:2016kko}
O.~J.~P. \'Eboli and M.~C. Gonzalez-Garcia, \emph{{Classifying the bosonic
  quartic couplings}},
  \href{http://dx.doi.org/10.1103/PhysRevD.93.093013}{\emph{Phys. Rev. D} {\bf
  93} (2016) 093013}, [\href{https://arxiv.org/abs/1604.03555}{{\tt
  1604.03555}}].

\bibitem{Guo:2015isa}
F.-K. Guo, P.~Ruiz-Femen\'\i{}a and J.~J. Sanz-Cillero, \emph{{One loop
  renormalization of the electroweak chiral Lagrangian with a light Higgs
  boson}}, \href{http://dx.doi.org/10.1103/PhysRevD.92.074005}{\emph{Phys. Rev.
  D} {\bf 92} (2015) 074005}, [\href{https://arxiv.org/abs/1506.04204}{{\tt
  1506.04204}}].

\bibitem{Alonso:2017tdy}
R.~Alonso, K.~Kanshin and S.~Saa, \emph{{Renormalization group evolution of
  Higgs effective field theory}},
  \href{http://dx.doi.org/10.1103/PhysRevD.97.035010}{\emph{Phys. Rev. D} {\bf
  97} (2018) 035010}, [\href{https://arxiv.org/abs/1710.06848}{{\tt
  1710.06848}}].

\bibitem{Buchalla:2017jlu}
G.~Buchalla, O.~Cata, A.~Celis, M.~Knecht and C.~Krause, \emph{{Complete
  One-Loop Renormalization of the Higgs-Electroweak Chiral Lagrangian}},
  \href{http://dx.doi.org/10.1016/j.nuclphysb.2018.01.009}{\emph{Nucl. Phys. B}
  {\bf 928} (2018) 93--106}, [\href{https://arxiv.org/abs/1710.06412}{{\tt
  1710.06412}}].

\bibitem{Buchalla:2020kdh}
G.~Buchalla, O.~Cat\`a, A.~Celis, M.~Knecht and C.~Krause,
  \emph{{Higgs-electroweak chiral Lagrangian: One-loop renormalization group
  equations}}, \href{http://dx.doi.org/10.1103/PhysRevD.104.076005}{\emph{Phys.
  Rev. D} {\bf 104} (2021) 076005},
  [\href{https://arxiv.org/abs/2004.11348}{{\tt 2004.11348}}].

\bibitem{Adler:1964um}
S.~L. Adler, \emph{{Consistency conditions on the strong interactions implied
  by a partially conserved axial vector current}},
  \href{http://dx.doi.org/10.1103/PhysRev.137.B1022}{\emph{Phys. Rev.} {\bf
  137} (1965) B1022--B1033}.

\bibitem{Adler:1965ga}
S.~L. Adler, \emph{{Consistency conditions on the strong interactions implied
  by a partially conserved axial-vector current. II}},
  \href{http://dx.doi.org/10.1103/PhysRev.139.B1638}{\emph{Phys. Rev.} {\bf
  139} (1965) B1638--B1643}.

\bibitem{Low:2014nga}
I.~Low, \emph{{Adler\textquoteright{}s zero and effective Lagrangians for
  nonlinearly realized symmetry}},
  \href{http://dx.doi.org/10.1103/PhysRevD.91.105017}{\emph{Phys. Rev. D} {\bf
  91} (2015) 105017}, [\href{https://arxiv.org/abs/1412.2145}{{\tt
  1412.2145}}].

\bibitem{Low:2014oga}
I.~Low, \emph{{Minimally symmetric Higgs boson}},
  \href{http://dx.doi.org/10.1103/PhysRevD.91.116005}{\emph{Phys. Rev. D} {\bf
  91} (2015) 116005}, [\href{https://arxiv.org/abs/1412.2146}{{\tt
  1412.2146}}].

\bibitem{Cheung:2014dqa}
C.~Cheung, K.~Kampf, J.~Novotny and J.~Trnka, \emph{{Effective Field Theories
  from Soft Limits of Scattering Amplitudes}},
  \href{http://dx.doi.org/10.1103/PhysRevLett.114.221602}{\emph{Phys. Rev.
  Lett.} {\bf 114} (2015) 221602}, [\href{https://arxiv.org/abs/1412.4095}{{\tt
  1412.4095}}].

\bibitem{Cheung:2015ota}
C.~Cheung, K.~Kampf, J.~Novotny, C.-H. Shen and J.~Trnka, \emph{{On-Shell
  Recursion Relations for Effective Field Theories}},
  \href{http://dx.doi.org/10.1103/PhysRevLett.116.041601}{\emph{Phys. Rev.
  Lett.} {\bf 116} (2016) 041601},
  [\href{https://arxiv.org/abs/1509.03309}{{\tt 1509.03309}}].

\bibitem{Low:2019ynd}
I.~Low and Z.~Yin, \emph{{Soft Bootstrap and Effective Field Theories}},
  \href{http://dx.doi.org/10.1007/JHEP11(2019)078}{\emph{JHEP} {\bf 11} (2019)
  078}, [\href{https://arxiv.org/abs/1904.12859}{{\tt 1904.12859}}].

\bibitem{Dai:2020cpk}
L.~Dai, I.~Low, T.~Mehen and A.~Mohapatra, \emph{{Operator Counting and Soft
  Blocks in Chiral Perturbation Theory}},
  \href{http://dx.doi.org/10.1103/PhysRevD.102.116011}{\emph{Phys. Rev. D} {\bf
  102} (2020) 116011}, [\href{https://arxiv.org/abs/2009.01819}{{\tt
  2009.01819}}].

\bibitem{Low:2022iim}
I.~Low, J.~Shu, M.-L. Xiao and Y.-H. Zheng, \emph{{Amplitude/Operator Basis in
  Chiral Perturbation Theory}},  \href{https://arxiv.org/abs/2209.00198}{{\tt
  2209.00198}}.

\bibitem{Grinstein:2007iv}
B.~Grinstein and M.~Trott, \emph{{A Higgs-Higgs bound state due to new physics
  at a TeV}}, \href{http://dx.doi.org/10.1103/PhysRevD.76.073002}{\emph{Phys.
  Rev. D} {\bf 76} (2007) 073002}, [\href{https://arxiv.org/abs/0704.1505}{{\tt
  0704.1505}}].

\bibitem{Buchalla:2013eza}
G.~Buchalla, O.~Cat\'a and C.~Krause, \emph{{On the Power Counting in Effective
  Field Theories}},
  \href{http://dx.doi.org/10.1016/j.physletb.2014.02.015}{\emph{Phys. Lett. B}
  {\bf 731} (2014) 80--86}, [\href{https://arxiv.org/abs/1312.5624}{{\tt
  1312.5624}}].

\bibitem{Gavela:2014vra}
M.~B. Gavela, J.~Gonzalez-Fraile, M.~C. Gonzalez-Garcia, L.~Merlo, S.~Rigolin
  and J.~Yepes, \emph{{CP violation with a dynamical Higgs}},
  \href{http://dx.doi.org/10.1007/JHEP10(2014)044}{\emph{JHEP} {\bf 10} (2014)
  044}, [\href{https://arxiv.org/abs/1406.6367}{{\tt 1406.6367}}].

\bibitem{Pich:2018ltt}
A.~Pich, \emph{{Effective Field Theory with Nambu-Goldstone Modes}},
  \href{https://arxiv.org/abs/1804.05664}{{\tt 1804.05664}}.

\bibitem{Gasser:1983yg}
J.~Gasser and H.~Leutwyler, \emph{{Chiral Perturbation Theory to One Loop}},
  \href{http://dx.doi.org/10.1016/0003-4916(84)90242-2}{\emph{Annals Phys.}
  {\bf 158} (1984) 142}.

\bibitem{Gasser:1984gg}
J.~Gasser and H.~Leutwyler, \emph{{Chiral Perturbation Theory: Expansions in
  the Mass of the Strange Quark}},
  \href{http://dx.doi.org/10.1016/0550-3213(85)90492-4}{\emph{Nucl. Phys. B}
  {\bf 250} (1985) 465--516}.

\bibitem{Buchalla:2016sop}
G.~Buchalla, O.~Cata, A.~Celis and C.~Krause, \emph{{Comment on ''Analysis of
  General Power Counting Rules in Effective Field Theory''}},
  \href{https://arxiv.org/abs/1603.03062}{{\tt 1603.03062}}.

\bibitem{Gavela:2016bzc}
B.~M. Gavela, E.~E. Jenkins, A.~V. Manohar and L.~Merlo, \emph{{Analysis of
  General Power Counting Rules in Effective Field Theory}},
  \href{http://dx.doi.org/10.1140/epjc/s10052-016-4332-1}{\emph{Eur. Phys. J.
  C} {\bf 76} (2016) 485}, [\href{https://arxiv.org/abs/1601.07551}{{\tt
  1601.07551}}].

\bibitem{Hirn:2005fr}
J.~Hirn and J.~Stern, \emph{{Lepton-number violation and right-handed neutrinos
  in Higgs-less effective theories}},
  \href{http://dx.doi.org/10.1103/PhysRevD.73.056001}{\emph{Phys. Rev. D} {\bf
  73} (2006) 056001}, [\href{https://arxiv.org/abs/hep-ph/0504277}{{\tt
  hep-ph/0504277}}].

\bibitem{Dobado:1989ax}
A.~Dobado and M.~J. Herrero, \emph{{Phenomenological Lagrangian Approach to the
  Symmetry Breaking Sector of the Standard Model}},
  \href{http://dx.doi.org/10.1016/0370-2693(89)90981-7}{\emph{Phys. Lett. B}
  {\bf 228} (1989) 495--502}.

\end{thebibliography}\endgroup
